\documentclass[12pt]{article}
\usepackage{a4wide}
\usepackage[ngerman,english]{babel}
\usepackage{amsmath, amssymb, latexsym}
\usepackage{graphicx}
\usepackage{bm}
\usepackage{hyperref}
\usepackage{color}

\begin{document}
\begin{titlepage}

\begin{center}
\vspace{2mm}

\par\end{center}

\begin{center}
\textbf{\Large{}Scattering of axial gravitational wave pulses by monopole black holes and QNMs}\textbf{ }
\par\end{center}

\begin{center}
Alexander Gu{\ss}mann
\\
 
\par\end{center}

\begin{center}
\textsl{\small{}Stanford Institute for Theoretical Physics, Stanford University}\\
\textsl{\small{} Stanford, California 94305, USA}\\
\texttt{\small{} gussmann@stanford.edu }\\
\par\end{center}{\small \par}

\vskip 0.3 cm
\begin{abstract}
We study scattering of short Gaussian pulses of axial gravitational waves by a black hole that has swallowed one or more global monopoles. We show how the response of the black hole to the impinging pulses depends both on the number of monopoles the black hole has swallowed and on the symmetry breaking scale of the model which gave rise to the monopoles. We determine the corresponding quasinormal modes that get excited by the impinging pulses and that get imprinted in the black hole's response to the pulses. Such modes are also expected to show up in various other dynamical processes such as the ringdown phase of a binary black hole merger in case at least one of the companion black holes of the binary has swallowed one or more global monopoles.
\end{abstract}
\end{titlepage}

\section{Introduction}

It has been known for a long time that black holes can be probed with short pulses of gravitational waves (see \cite{Vishveshwara:1970zz} for the pioneering work): When a gravitational wave pulse with a width that is comparable or less than the size of a black hole is scattered by the black hole, the impinging pulse excites the black hole's (tensor) quasinormal modes which get then imprinted in the scattered outgoing wave pulse \cite{Vishveshwara:1970zz, Vishveshwara:1996jgz, Andersson:1995zk}. Since quasinormal modes only depend on the parameters that characterize the black hole \cite{Press:1971wr}, this implies that such scattered pulses can be useful both to determine the values of parameters of a given black hole and to investigate whether or not a given black hole is completely characterized by the parameters that a theoretical model predicts.

Although many aspects about (tensor) quasinormal modes and the scattering of gravitational wave pulses by black holes or other compact objects have been studied in the literature (see e.g. \cite{Kokkotas:1999bd, Berti:2009kk, Konoplya:2011qq} for reviews on quasinormal modes and \cite{Vishveshwara:1970zz, Andersson:1995zk, Andersson:1996pn, Allen:1997xj} for some works on scattering of wave pulses), to our knowledge, a theoretical analysis for the case of global monopole black holes is still missing.\footnote{However, some related works exist. Scattering and absorption of plane (scalar) waves by such black holes was studied in \cite{Hai:2013ara, Anacleto:2017kmg, Pitelli:2017bgx}. Scalar and spinor quasinormal modes in $f(R)$ gravity were investigated in \cite{Graca:2015jea}. Quasinormal modes in the presence of quintessence were determined in \cite{Xi:2008ce}.}
Such black holes can be formed when global monopoles that were produced in phase transitions in the early universe get swallowed by black holes or when overdense regions of matter that contain such monopoles collapse to black holes.
 
In this work we shall both determine tensor quasinormal modes of spherically symmetric monopole black holes and study the quasinormal mode contribution to the scattering of Gaussian axial gravitational wave pulses by global monopole black holes. The aim is not to provide precise quantitative predictions but rather to show the qualitative behavior. Since quasinormal modes of a black hole not only show up in scattering setups but also in various other dynamical processes such as the ringdown phase of black hole binary mergers, our results also have implications for the understanding of such different processes whenever a global monopole black hole is involved.

The paper is organized as follows: In section \ref{section1}, we review some aspects about global monopoles and global monopole black holes. In section \ref{section2}, we determine tensor quasinormal modes of such black holes and investigate how they depend both on the symmetry breaking scale of the model that gave rise to the monopoles and on the number of monopoles that are inside of the black holes. In section \ref{section3}, we study the scattering processes of the Gaussian axial gravitational wave pulses. In section \ref{section4}, we conclude with a discussion where we comment on the question whether or not the effects that we found in our theoretical analysis can have observable manifestations.

We use units in which $c = \hbar = 1$. For the metric we use the signature $(+, -, -, -)$. Although many numerical techniques exist, we mostly use known (semi-)analytical methods that we apply to the case of axial gravitational wave pulses scattered by monopole black holes. This requires several approximations to be made. As mentioned above, our results should therefore not be taken as precise quantitative predictions but are rather meant to show the correct qualitative behavior. The hope is that the analytical analysis provides a better insight into the underlying physics when compared to a pure numerical analysis.

\section{Global monopole black holes} \label{section1}

\subsection{Global monopoles} \label{section11}

Global monopoles are topological defects that arise in models with a global symmetry $A$ that is spontaneously broken down to a symmetry $B$ in such a way that the second homotopy group $\pi_2(A/B)$ is nontrivial. The simplest such model with a global symmetry $O(3)$ that is spontaneously broken down to $O(2)$, first mentioned by Polyakov \cite{Pol:1974}, is described by the Lagrangian density
\begin{equation}
\mathcal{L} = \frac{1}{2} \partial_\mu \phi^a \partial^\mu \phi^a - \frac{\lambda}{4} \left(\phi^a \phi^a - v^2\right)^2 \, .
\label{lag}
\end{equation}
Here $\phi^a$ is a scalar triplet field ($a = 1,2,3$), $\lambda$ a self-coupling constant and $v$ the vacuum expectation value. Making a hedgehog ansatz in spherical coordinates,
\begin{equation}
\phi^a = v h(r) n^a \, ,
\label{field}
\end{equation}
where $n^a$ is a unit vector in radial direction, and using boundary conditions $h(0) = 0$ and $h(\infty) = \pm 1$, one can find classical solutions of the corresponding equations of motion for $h(r)$ which have a non-zero topological charge \cite{Pol:1974} 
\begin{equation}
Q = \frac{1}{4 \pi v^3} \oint_{S^2} \epsilon_{abc} \phi^a \partial_i \phi^b \partial_j \phi^c dx^i \wedge dx^j \, .
\end{equation}
Solutions with $Q = 1$ (corresponding to $h(\infty) =1$) are often referred to as ``global monopoles", whereas solutions with $Q = -1$ ($h(\infty) = -1$) are known as ``global antimonopoles".

Global monopoles have an important property: Outside of the core of a monopole, where the solution-function $h(r) \approx 1$, the temporal-temporal component of the energy momentum tensor of the monopole takes the form $T^0_0 = \frac{v^2}{r^2}$. Therefore, the total energy $E$ of one global monopole asymptotically scales with radial distance $r$ as 
\begin{equation}
E = 4 \pi \int_0^r T_0^0 r^2 dr \sim M_c + 4 \pi v^2 r
\label{energy}
\end{equation}
and is thus linearly divergent. (Here $M_c \sim \frac{v}{\sqrt{\lambda}}$ is the mass of the monopole core \cite{Barriola:1989hx}.) This implies that, if one wants to consider finite energy objects in an infinite space, one cannot consider single isolated global monopoles. One could argue that there is no need for considering finite energy objects as long as the total number of global monopoles in our universe (with infinite total energy) is small enough such that their (finite) energy density does not overclose the universe. If one however wants to consider finite energy objects, there are two possible ways, already pointed out by Polyakov \cite{Pol:1974}, to obtain finite energy configurations from (\ref{lag}): The first possibility is to gauge the symmetry in (\ref{lag}). This gives rise to local finite energy magnetic monopoles (today well-known as 't Hooft Polyakov monopoles \cite{Pol:1974, tHooft:1974kcl}). In this work, we will however not consider such local magnetic monopoles. The second possibility is to consider (a network of) global monopole antimonopole pairs instead of single global monopoles or, in other words, an equal number of global monopoles and global antimonopoles. In this second case the divergences of the energies of the monopole and antimonopole of a pair cancel. The energy of one pair is then $\sim 2 M_c + 4 \pi v^2 R$, where $R$ is the distance between the monopole and the antimonopole of the pair. This leads to an attractive force $F$ between the global monopole and the global antimonopole of a pair that does not depend on the distance between them, $F = \partial_R E \sim 4 \pi v^2$.

 For the theoretical analysis of this work, it is not important whether or not there is an equal number of global monopoles and global antimonopoles in our universe. We will however come back to this point and to the question whether global monopoles exist at all in our universe in the discussion section \ref{section4} where we comment on possible formation mechanisms and possible observable manifestations. 

Although this has been a matter of debate in the literature \cite{Goldhaber:1989na, Rhie:1990kc, Perivolaropoulos:1991du}, global monopoles seem to be dynamically stable if they can freely move \cite{Achucarro:2000td}. In particular, numerical simulations indicate that there is no instability that causes the field of a freely moving monopole antimonopole pair to collapse with no energy cost to a string \cite{Bennett:1990xy}.

\subsection{Black holes} \label{section12}

If existent in our universe, global monopoles and global antimonopoles can happen to be inside of a black hole either when they are swallowed by the black hole or when an overdense region of matter that contains global monopoles and/or global antimonopoles collapses to a black hole. We will comment on the possible formation mechanisms in some detail in the discussion section \ref{section4}. In what follows we shall recall how these objects (global monopoles inside of black holes) are described theoretically.

When (\ref{lag}) is minimally coupled to Einstein gravity, one can find gravitating solutions of the Einstein field equations, $G_{\mu \nu} = \frac{1}{M_P^2}T_{\mu \nu}$, both without event horizon (``gravitating monopoles") and with event horizon (``monopole black holes") \cite{Barriola:1989hx}: Far away from the monopole core where $h(r) \approx 1$, the energy momentum tensor for one global monopole takes the form
\begin{equation}
T^\mu_\nu = {\rm diag} \left(\frac{v^2}{r^2}, \frac{v^2}{r^2},0,0\right) \, ,
\label{emo}
\end{equation}
which implies anisotropic stress, $p_r \neq p_t$, (with $p_r \equiv -T_r^r$ the radial pressure and $p_t \equiv -T_\theta^\theta$ the tangential pressure) and an energy density $\rho \equiv T_0^0$ that is equal to minus the radial pressure $p_r$, $\rho = -p_r$. A solution of the Einstein field equations with this energy momentum tensor (\ref{emo}) was found in \cite{Barriola:1989hx} and is given by the line element
\begin{equation}
ds^2 = A(r) dt^2 - A(r)^{-1} dr^2 - r^2\left(d \theta^2 + {\rm sin}^2 \theta d \phi^2\right) \, ,
\label{metric}
\end{equation}
with
\begin{equation}
A(r) = 1 - \frac{v^2}{M_P^2}-\frac{2M}{M_P^2r} \, .
\label{A2}
\end{equation}
Here $M_P$ is the reduced Planck mass and $M$ is a constant of integration. As indicated above, one often distinguishes two cases. The first case is the case where $M$ is set only by the monopole energy momentum tensor, $A(r) \equiv 1 - \frac{1}{M_P^2r}\int T_0^0 r^2 dr$. In that case $M$ takes the value of $M_c (8 \pi)^{-1}$. Such configurations are referred to as ``gravitating monopoles".\footnote{It can be shown that $M_c$ is a negative quantity leading to a repulsive Newton potential of the global monopole core \cite{Harari:1990cz}. Non-minimal couplings to gravity \cite{Nucamendi:2000is, Nucamendi:2000ch} or an additional unbroken $U(1)$ subgroup \cite{Dvali:2000ty, Spinelly:2002mt, Achucarro:2002gg, BezerradeMello:2003yt} can however give similar gravitating monopole solutions with attractive Newton potential of the core. Except of the small (repulsive) potential of the monopole core, there is no gravitational force that a gravitating global monopole exerts on the matter around it.} The second case is the case where $M$ is not set by $M_c$ but takes large positive values such that the configuration (\ref{metric}) has an event horizon and describes a large black hole (``monopole black hole"). The asymptotic form of the monopole black hole solution that we gave above (\ref{metric}), (\ref{emom}) is a good approximation for large monopole black holes (with the size of the black hole much larger than the core of the monopole), precise numerical solutions which are particularly important for intermediate regimes however also exist \cite{Liebling:1999bb, Maison:1999ke, Nucamendi:2000af, Tamaki:2003kv}. The stability of these objects was discussed in \cite{Watabe:2004tq}. In what follows we shall use the asymptotic form of the global monopole black holes (\ref{emo}), (\ref{metric}), (\ref{A2}) generalized to the case where there are $N$ global monopoles inside of the black hole (or $N_a$ global antimonopoles and $N+N_a$ global monopoles):
\begin{equation}
T^\mu_\nu = {\rm diag}\left(N\frac{v^2}{r^2}, N\frac{v^2}{r^2},0,0\right) \, ,
\label{emom}
\end{equation}
\begin{equation}
A(r) = 1 - \frac{Nv^2}{M_P^2}-\frac{2M}{M_P^2r} \, .
\label{Ar}
\end{equation}

\section{Quasinormal modes of monopole black holes} \label{section2}

As other kinds of black holes, global monopole black holes possess characteristic quasinormal modes (see e.g. \cite{Kokkotas:1999bd, Berti:2009kk, Konoplya:2011qq} for reviews on quasinormal modes). In this section, we shall determine the axial tensor quasinormal modes of monopole black holes with the help of semianalytical methods. We first derive the linearized Einstein field equations for the axial modes (odd parity modes) in Regge-Wheeler gauge. They reduce to a single second order differential equation for the perturbations that, as one can expect, goes to the Regge-Wheeler equation \cite{Regge:1957td} in the limit $v \rightarrow 0$. In the next section we shall discuss one way of how to excite these quasinormal modes in a dynamical process.

\subsection{Linearized Einstein field equations} \label{section21}

\begin{figure}
\includegraphics[scale=0.65]{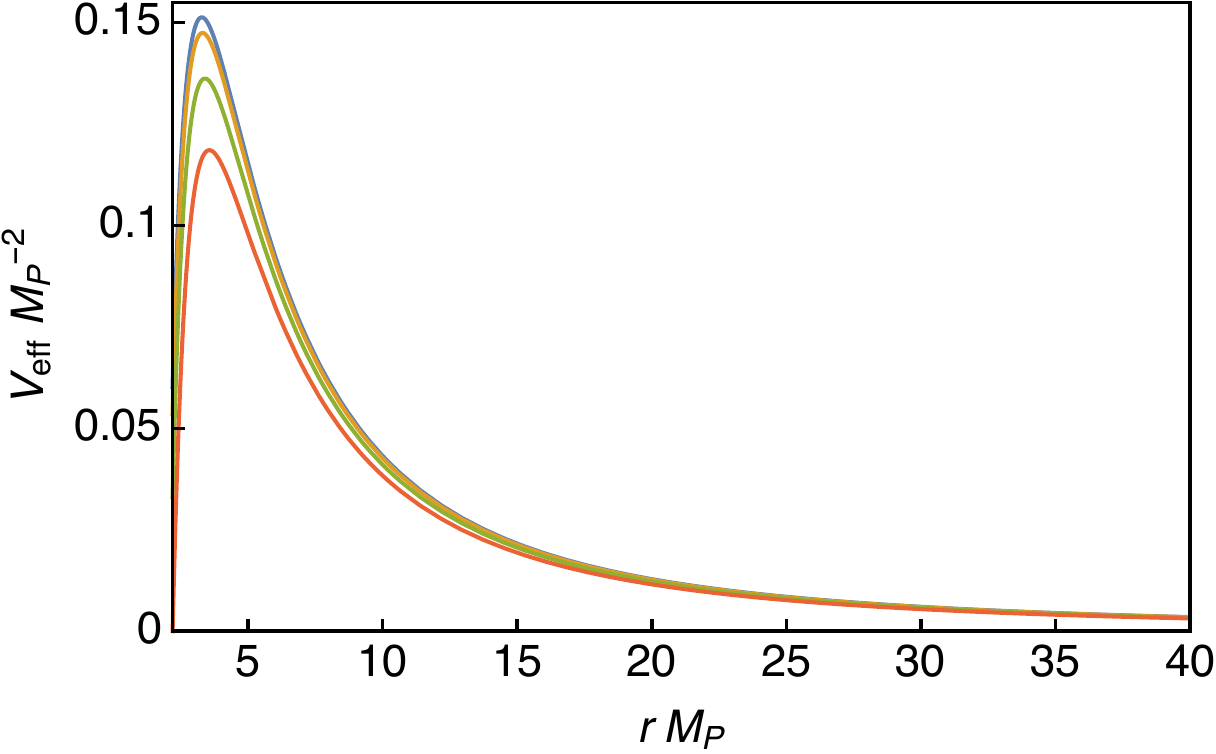}
\includegraphics[scale=0.65]{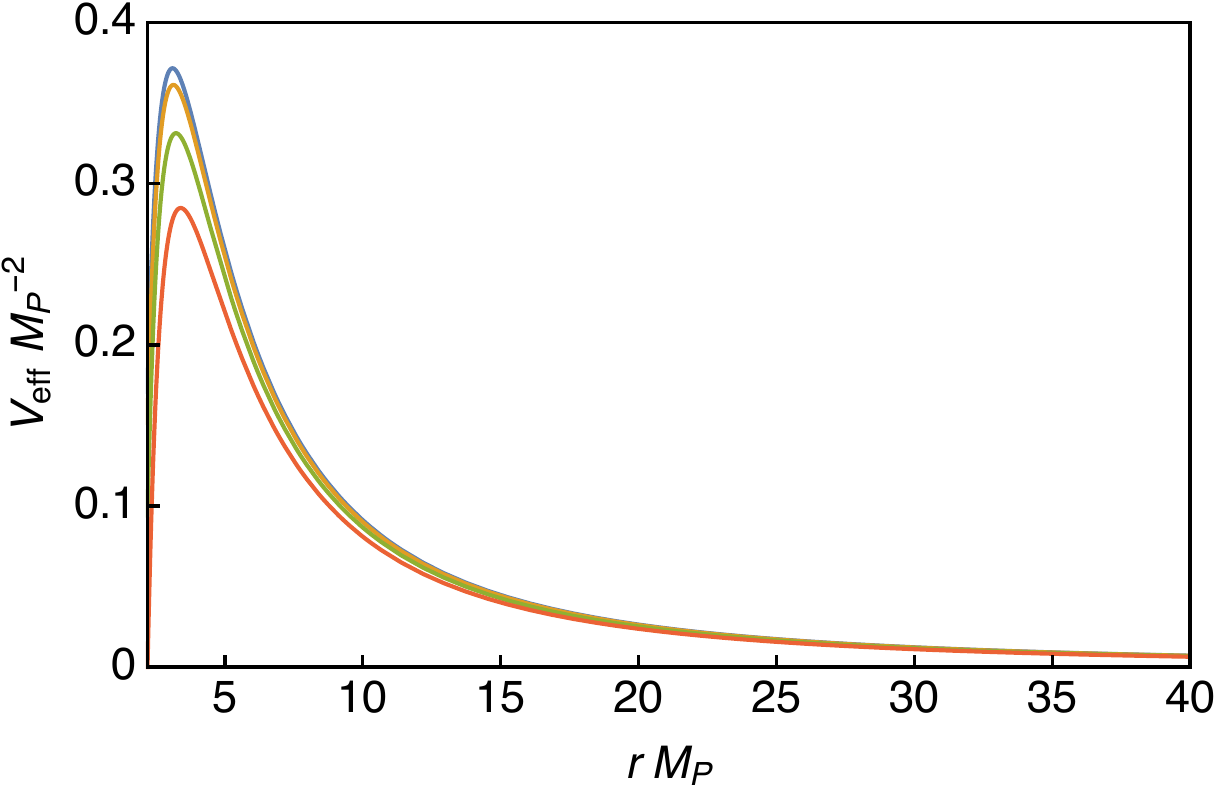}
\caption{The effective potential (\ref{effective}) for $l=2$ (left) and $l=3$ (right) for the case $M=M_P$. The different colors represent different values of $Nv^2$: The blue line is the Schwarzschild case $Nv^2 = 0$, the yellow line is the case $Nv^2 = 0.01 M_P^2$, the green line is $Nv^2 = 0.04 M_P^2$ and the red line is $Nv^2 = 0.09 M_P^2$.}
\label{figure1}
\end{figure}

In Regge-Wheeler gauge the perturbed line element can be written as \cite{Regge:1957td}
\begin{equation}
ds^2 = ds_0^2 +ds_{{\rm (polar)}}^2 + ds_{{\rm (axial)}}^2 \, ,
\end{equation}
where $ds_0^2$ is the line element of the background metric (\ref{metric}) with $A(r)$ as defined in (\ref{Ar}) and
\begin{equation}
ds_{{\rm (polar)}}^2 \equiv \left(H_0 A(r) dt^2 + 2H_1 drdt + H_2 A(r)^{-1} dr^2+ r^2 K d \theta^2 + r^2 K {\rm sin}^2 \theta  d \phi^2\right) e^{im\phi}P_l({\rm cos} \theta) \, ,
\label{polar}
\end{equation}
\begin{equation}
ds_{{\rm (axial)}}^2 \equiv 2h_0e^{im \phi} {\rm sin} \theta \partial_\theta P_l({\rm cos} \theta) dt d\phi + 2h_1e^{im\phi} {\rm sin} \theta \partial_\theta P_l({\rm cos} \theta) dr d\phi \, .
\label{axial}
\end{equation}
Here $P_l$ are the standard Legendre polynomials of order $l$. $H_0$, $H_1$, $H_2$, $K$, $h_0$ and $h_1$ are functions of only the temporal coordinate $t$ and the radial coordinate $r$. These expressions can be used to determine the perturbations of the Einstein tensor $\delta G_{\mu \nu}$.\footnote{We have determined the non-vanishing components for the axial modes of $\delta G_{\mu \nu}$ and provide them in the appendix \ref{appendixa}.} The perturbations of the energy momentum tensor are given by 
\begin{equation}
\delta T_{\mu \nu} = \delta \left(\partial_\mu \phi^a\right) \partial_\nu \phi^a + \partial_\mu \phi^a \delta \left(\partial_\nu \phi^a\right) - h_{\mu \nu} \mathcal{L} - g_{\mu \nu} \delta \mathcal{L} \, ,
\end{equation}
where $g_{\mu \nu}$ are the components of the background metric (\ref{metric}) and $h_{\mu \nu}$ are the components of the metric perturbations (\ref{polar}), (\ref{axial}).
We restrict our studies here to the axial modes. Since perturbations of the scalar field are polar, $\delta \phi^{{\rm (axial)}} = 0$, the axial modes of the perturbations of the energy momentum tensor $\delta T_{\mu \nu}^{{\rm (axial)}}$ are, in our case of the global monopole with background energy momentum tensor (\ref{emom}), given by
\begin{equation}
\delta T_{\mu \nu}^{{\rm (axial)}} = h_{\mu \nu}^{{\rm (axial)}} N \frac{v^2}{r^2} \, .
\end{equation}
Here $h_{\mu \nu}^{{\rm (axial)}}$ are the components of the axial metric perturbations (\ref{axial}). The non-vanishing components of the perturbed Einstein field equations for the axial modes, $\delta G_{\mu \nu}^{{\rm (axial)}} = \frac{1}{M_P^2} \delta T_{\mu \nu}^{{\rm (axial)}}$, can for each frequency component be written as
\begin{equation}
A(r) \partial_r \hat{h}_1 + i w A(r)^{-1} \hat{h}_0 + \frac{2}{r^2}\frac{M}{M_P^2} \hat{h}_1 = 0 \, ,
\label{eq1}
\end{equation}
\begin{equation}
\frac{1}{r} A(r)^{-1} \left(2 i w \hat{h}_0 + r w^2 \hat{h}_1 -i w r \partial_r \hat{h}_0\right) + \frac{2}{r^2} \hat{h}_1 - \frac{1}{r^2}\left(l\left(l+1\right)\right) \hat{h}_1 - \frac{2N}{M_P^2} \frac{v^2}{r^2} \hat{h}_1 = 0 \, ,
\label{eq2}
\end{equation}
\begin{equation}
A(r) \partial_r^2 \hat{h}_0 + iw A(r) \partial_r \hat{h}_1 + \frac{2iw}{r}A(r) \hat{h}_1 - \frac{1}{r^2} \left(l\left(l+1\right)\right) \hat{h}_0 + \frac{4 M}{M_P^2 r^3} \hat{h}_0 = 0 \, ,
\label{eq3}
\end{equation}
where $A(r)$ is defined as in (\ref{Ar}) and $\hat{h}_0$ and $\hat{h}_1$ are the Fourier transforms of $h_0$ and $h_1$ and thus functions of the radial coordinate $r$ and of the frequency $w$. Here (\ref{eq1}) is the $\theta \phi$ component of the perturbed Einstein field equations, (\ref{eq2}) the $r \phi$ component and (\ref{eq3}) the $t \phi$ component. The last equation (\ref{eq3}) is a consequence of (\ref{eq1}) and (\ref{eq2}). Similarly as for example in the case of a Schwarzschild black holes (see e.g. \cite{Regge:1957td}) and in the case of a static perfect fluid star (see e.g. \cite{Chandrasekhar:1991fi}), one can eliminate $\hat{h}_0$ from (\ref{eq2}) by using (\ref{eq1}). Then (\ref{eq2}) reduces to the second order differential equation
\begin{equation}
A(r)\partial_r\left(A(r)\partial_r \hat{\Psi}\right)+\left(w^2-V_{{\rm eff}}(r)\right)\hat{\Psi} = 0 \, ,
\label{schr}
\end{equation}
where $\hat{\Psi} \equiv A(r) r^{-1} \hat{h}_1$ and
\begin{equation}
V_{{\rm eff}}(r) \equiv A(r)\left(\frac{l\left(l+1\right)}{r^2} - \frac{6M}{M_P^2 r^3}\right) \, .
\label{effective}
\end{equation}
The result (\ref{schr}) is a Schroedinger equation with the effective potential $V_{{\rm eff}}$. Scattering of axial gravitational waves can therefore be studied by using techniques from one-dimensional quantum mechanical scattering theory. In the limit $v \rightarrow 0$, (\ref{schr}) goes to the Regge-Wheeler equation \cite{Regge:1957td}. We plot the effective potentials for $l=2$, $l=3$ and several values of $Nv^2$ in figure \ref{figure1}.

\subsection{Quasinormal modes} \label{section22}

The wave equation (\ref{schr}) can be used to determine the corresponding quasinormal modes. By definition these are the eigenfunctions of (\ref{schr}) which are subject to the boundary conditions
\begin{equation}
\hat{\Psi} \sim e^{-iwr^*}, r^* \rightarrow - \infty \, ,
\label{in}
\end{equation}
\begin{equation}
\hat{\Psi} \sim e^{iwr^*}, r^* \rightarrow \infty \, .
\label{out}
\end{equation}
Here $r^*$ is the tortoise coordinate that is defined by $\partial_{r^*} \equiv A(r) \partial_r$.\footnote{Note that it follows from (\ref{effective}) and the definition of $r^*$ that $V_{{\rm eff}} \rightarrow 0$ for $r^* \rightarrow \pm \infty$.}
There exist several techniques to determine the corresponding complex eigenfrequencies. In this work, we shall use the WKB method that yields the quasinormal mode frequencies to a good approximation \cite{Schutz:1985km}, see e.g.  \cite{Konoplya:2019hlu} for a recent summary of the method. We determine quasinormal mode frequencies for $l=2$ and $l=3$ modes both by using the approximation carried to third order beyond eikonal approximation (both for the $l=2$ modes and for the $l=3$ modes) \cite{Iyer:1986np} and by using the approximation carried out to eighth order (for the $l=2$ modes) and to tenth order (for the $l=3$ modes) combined with Pad\'e approximations \cite{Matyjasek:2017psv, Matyjasek:2019eeu, Hatsuda:2019eoj}. The latter is the state-of-art technique to obtain very accurate results with the WKB method (see e.g. \cite{Matyjasek:2017psv, Matyjasek:2019eeu, Konoplya:2019hlu}). 

When carrying out the approximation to third order, the quasinormal mode frequencies $w_n$ ($n=0,1,2, ...$) are given by \cite{Iyer:1986np, Kokkotas:1988fm}
\begin{equation}
w_n^2 = V_0 + \sqrt{-2V_0^{(2)}}\Lambda(n) - i \left(n + \frac{1}{2}\right)\sqrt{-2V_0^{(2)}}\left(1 + \Omega(n)\right) \, ,
\end{equation}
where
\begin{equation}
\Lambda(n) \equiv \frac{1}{8 \sqrt{-2V_0^{(2)}}}\left(\frac{V_0^{(4)}}{V_0^{(2)}}\left(\frac{1}{4} + \left(n+\frac{1}{2}\right)^2\right) - \frac{1}{36}\left(\frac{V_0^{(3)}}{V_0^{(2)}}\right)^2 \left(7+60 \left(n+\frac{1}{2}\right)^2\right)\right) \, ,
\end{equation}
\begin{equation}
\begin{split}
\Omega(n) \equiv& \frac{-1}{2V_0^{(2)}}{\Bigg{(}}\frac{5}{6912}\left(\frac{V_0^{(3)}}{V_0^{(2)}}\right)^4\left(77+188 \left(n+\frac{1}{2}\right)^2\right)  - \frac{1}{384}\left(\frac{(V_0^{(3)})^2V_0^{(4)}}{(V_0^{(2)})^3}\right) \\
& \Bigg(51 +100 \left(n + \frac{1}{2}\right)^2\Bigg) + \frac{1}{2304}\left(\frac{V_0^{(4)}}{V_0^{(2)}}\right)^2 \left(67+68 \left(n + \frac{1}{2}\right)^2\right) \\
& + \frac{1}{288} \left(\frac{V_0^{(3)}V_0^{(5)}}{(V_0^{(2)})^2}\right)\left(19 + 28 \left(n + \frac{1}{2}\right)^2\right) - \frac{1}{288}\left(\frac{V_0^{(6)}}{V_0^{(2)}}\right)\left(5+4 \left(n+\frac{1}{2}\right)^2\right) {\Bigg{)}} \, .
\end{split}
\end{equation}
Here $V_0^{(k)}$ stands for the k-th derivative $\frac{d^kV_{{\rm eff}}}{dr^{*k}}$ evaluated at the peak $r^*_0$ of the effective potential (\ref{effective}). When carrying out the approximation to higher order, in particular in the case when the approximation is combined with the Pad\'e approximation (as proposed in \cite{Matyjasek:2017psv, Matyjasek:2019eeu, Hatsuda:2019eoj}), the expressions can be found for example in the Mathematica code which is publicly available to download at \cite{code} (see also \cite{Konoplya:2003ii} for the expressions up to sixth order).

Using these expressions, we have determined the dimensionless quasinormal mode frequencies $w_n M_P^{-1}$ for $l = 2, 3$, $n = 0, 1, 2, 3$, $M=M_P$ and several values of $Nv^2$. We list our results both for the third order and for the eighth order (tenth order respectively) in table \ref{table1}. In the case of the eighth order (tenth order) results we have used the averaging technique over Pad\'e approximations that is presented in \cite{Konoplya:2019hlu, code} for each quasinormal mode frequency. The results for $Nv^2 = 0$ correspond to the Schwarzschild case, they are very close to the exact numerical results for Schwarzschild black holes which were determined in \cite{Leaver:1985ax, Andersson947}. As one can see from the results in table \ref{table1}, once $Nv^2$ increases (while $l$ and $n$ are kept fixed), the real part of the quasinormal mode frequencies decreases whereas the imaginary part increases. This qualitative behavior of the modes $w_n$ is both visible from the third order values as well as from the more precise higher order values.

\begin{table}[p]
\centering
\caption{Axial tensor quasinormal modes $w_n$ for the case $M=M_P$ determined both by using the WKB expansion carried out to third order beyond eikonal approximation and by using the WKB expansion carried out to eighth order (for the $l=2$ modes) and to tenth order (for the $l=3$ modes). In the latter case averaging over Pad\'e approximations as discussed in \cite{Konoplya:2019hlu, code} is used. The corresponding values for $\eta(w_n)$ and $\frac{\partial \gamma}{\partial w}\big|_{w = w_n}$ are determined by using the (less accurate) third order values of $w_n$ for the purpose to derive the qualitative behavior of the response of a black hole to an impinging wave pulse.}
\begin{tabular}{c c c c c c c}
\hline\hline
$l$ & $n$ & $\frac{Nv^2}{M_P^2}$ & $\frac{w_n}{M_P}$ {\tiny(3rd order WKB)} & $\frac{w_n}{M_P}$ {\tiny(8th/10th order WKB)} & $\eta(w_n)$ & $M_P \frac{\partial \gamma}{\partial w}\big|_{w=w_n}$ \\
\hline
2 & 0 & 0 & $0.3732 - i 0.0892$ & $0.3737 - i 0.0890$ & $-3.9364 - i 0.5679$ & $-1.4906+i17.3224$ \\
2 & 0 & 0.01 & $0.3686- i 0.0875$ & $0.3691 - i 0.0873$ & $-3.9526-i0.5701$ & $-1.5056+i17.6670$ \\
2 & 0 & 0.04 & $0.3549 - i 0.0825$ & $0.3553 - i 0.0823$ & $-4.0029-i0.5767$ & $-1.5528+i18.7642$ \\
2 & 0 & 0.09 & $0.3320 - i 0.0744$ & $0.3324 - i 0.0742$ & $-4.0927-i0.5878$ & $-1.6399+i20.8343$ \\
2 & 1 & 0 & $0.3460 - i 0.2749$ & $0.3468 - i 0.2737$ & $-4.8442 - i 0.9015$ & $-3.0148+i15.8116$ \\
2 & 1 & 0.01 & $0.3422 - i 0.2695$ & $0.3430 - i 0.2684$ & $-4.8599-i0.9076$ & $-3.0566+i16.1422$ \\
2 & 1 & 0.04 & $0.3307 - i 0.2537$ & $0.3314 - i 0.2528$ & $-4.9086-i0.9262$ & $-3.1881+i17.1965$ \\
2 & 1 & 0.09 & $0.3111 - i 0.2284$ & $0.3118 - i 0.2278$ & $-4.9958-i0.9577$ & $-3.4308+i19.1889$ \\
2 & 2 & 0 & $0.3029 - i 0.4711$ & $0.3002 - i 0.4766$ & $-5.8780 - i 0.8873$ & $-3.6587+i14.8384$ \\
2 & 2 & 0.01 & $0.3003 - i 0.4618$ & $0.2976 - i 0.4672$ & $-5.8925 - i 0.8973$ & $-3.7111+i15.1539$ \\
2 & 2 & 0.04 & $0.2920 - i 0.4344$ & $0.2896 - i 0.4395$ & $-5.9379-i0.9275$ & $-3.8766+i16.1598$ \\
2 & 2 & 0.09 & $0.2776 - i 0.3907$ & $0.2761 - i 0.3947$ & $-6.0193-i0.9788$ & $-4.1845+i18.0616$ \\
2 & 3 & 0 & $0.2475-i0.6729$ & $0.2427 - i 0.7076$ & $-6.9994-i0.6926$ & $-4.2870+i14.1089$ \\
2 & 3 & 0.01 & $0.2463 - i 0.6595$ & $0.2419 - i 0.6926$ & $-7.0127 - i 0.7066$ & $-4.3478+i14.4173$ \\
2 & 3 & 0.04 & $0.2424 - i 0.6203$ & $0.2391 - i 0.6492$ & $-7.0542-i0.7488$ & $-4.5396+i15.4007$ \\
2 & 3 & 0.09 & $0.2347-i0.5576$ & $0.2326 - i 0.5810$ & $-7.1294-i0.8201$ & $-4.8947+i17.2588$ \\
3 & 0 & 0 & $0.5993 - i 0.0927$ & $0.5994 - i 0.0927$ & $-5.3151-i0.6919$ & $-0.8220+i16.8429$ \\
3 & 0 & 0.01 & $0.5910 - i 0.0909$ & $0.5912 - i 0.0909$ & $-5.3393-i0.6934$ & $-0.8325+i17.1794$ \\
3 & 0 & 0.04 & $0.5663 - i 0.0856$ & $0.5664 - i 0.0856$ & $-5.4141-i0.6981$ & $-0.8658+i18.2525$ \\
3 & 0 & 0.09 & $0.5256 - i 0.0771$ & $0.5258 - i 0.0770$ & $-5.5471-i0.7061$ & $-0.9274+i20.2810$ \\
3 & 1 & 0 & $0.5824 - i 0.2814$ & $0.5826 - i 0.2813$ & $-6.1904-i1.2600$ & $-2.0103+i16.1220$ \\
3 & 1 & 0.01 & $0.5745 - i 0.2759$ & $0.5748 - i 0.2758$ & $-6.2141-i1.2646$ & $-2.0400+i16.4515$ \\
3 & 1 & 0.04 & $0.5511 - i 0.2596$ & $0.5513 - i 0.2595$ & $-6.2876-i1.2784$ & $-2.1339+i17.5026$ \\
3 & 1 & 0.09 & $0.5124 - i 0.2336$ & $0.5126 - i 0.2335$ & $-6.4183-i1.3021$ & $-2.3080+i19.4913$ \\
3 & 2 & 0 & $0.5532 - i 0.4767$ & $0.5517 - i 0.4791$ & $-7.1576-i1.4714$ & $-2.6138+i15.3905$ \\
3 & 2 & 0.01 & $0.5461 - i 0.4673$ & $0.5446 - i 0.4696$ & $-7.1805-i1.4788$ & $-2.6560+i15.7086$ \\
3 & 2 & 0.04 & $0.5247 - i 0.4395$ & $0.5234 - i 0.4416$ & $-7.2514-i1.5014$ & $-2.7900+i16.7237$ \\
3 & 2 & 0.09 & $0.4893 - i 0.3952$ & $0.4882 - i 0.3969$ & $-7.3776-i1.5404$ & $-3.0398+i18.6463$ \\
3 & 3 & 0 & $0.5157 - i 0.6774$ & $0.5120 - i 0.6904$ & $-8.2036-i1.4962$ & $-3.0162+i14.8870$ \\
3 & 3 & 0.01 & $0.5095 - i 0.6640$ & $0.5058 - i 0.6765$ & $-8.2254-i1.5064$ & $-3.0652+i15.1958$ \\
3 & 3 & 0.04 & $0.4908 - i 0.6245$ & $0.4873 - i 0.6356$ & $-8.2930-i1.5372$ & $-3.2204+i16.1812$ \\
3 & 3 & 0.09 & $0.4596 - i 0.5613$ & $0.4563 - i 0.5705$ & $-8.4137-i1.5904$ & $-3.5110+i18.0475$ \\
\hline
\end{tabular}
\label{table1}
\end{table}

\section{Scattering of axial gravitational wave pulses by global monopole black holes} \label{section3}

The quasinormal modes of a black hole get excited in many dynamical processes. One example of such a process is the scattering of a short wave pulse by the black hole: An impinging wave pulse with a width less than or comparable to the size of the black hole excites the quasinormal modes of the black hole which then get imprinted in the black hole's response to the pulse \cite{Vishveshwara:1970zz, Vishveshwara:1996jgz, Andersson:1995zk, Andersson:1996cm, Frolov:1998wf, Andersson:2000tf}.

We shall study scattering of a Gaussian axial gravitational wave pulse by a monopole black hole which at some initial point of time $t_0$ takes the form
\begin{equation}
\Psi(t_0, r^*) = a e^{-b \left(r^*-x\right)^2} \, ,
\nonumber
\end{equation}
\begin{equation}
\frac{\partial \Psi}{\partial t} \bigg|_{t=t_0} = 0 \, .
\label{initial}
\end{equation}
Here $a$ and $b$ are constants and $x$ is a point far away from the black hole at which the initial static Gaussian pulse is centered. The function $\Psi$ is the Fourier transform of the function $\hat{\Psi}$ that was introduced in (\ref{schr}). This kind of initial data was also used in other works (see e.g. \cite{Vishveshwara:1970zz, Andersson:1995zk}). We consider the Cauchy problem (\ref{schr}), (\ref{initial}) and want to find the outgoing wave solutions for large $r$ in order to investigate how the response of the black hole to an impinging pulse with initial data (\ref{initial}) depends on $Nv^2$. We focus on the quasinormal mode contribution to the solutions. This is the most characteristic contribution in such scattering setups for impinging pulses with a width smaller than or comparable to the size of the black hole. It dominates the response of the black hole at all but very early and very late times \cite{Leaver:1986gd, Andersson:1996cm}.\footnote{At very late and very early times, contributions from low frequencies (for very late times) and high frequencies (for very early times) dominate over the quasinormal mode contribution (see e.g. \cite{Andersson:1996cm}). The late time contributions lead to a power-law tail of the outgoing wave \cite{Price:1971fb, Price:1972pw, Cunningham:1978zfa, Leaver:1986gd, Andersson:1996cm}. Following for example the procedure in \cite{Andersson:1996cm} for a Schwarzschild black hole, it is straightforward to get the high and low frequency contributions also for the monopole black hole case. In this work, we only study the quasinormal mode contribution.} The hope is that the study of this scattering process will give us a good understanding of the effects that global monopoles which were swallowed by a black hole can have on dynamical processes in which quasinormal modes of the black hole show up.

Let us first mention that in the setup which we are considering there is an event horizon at $r^* \rightarrow - \infty$. Therefore, we demand the absence of outgoing waves when $r^* \rightarrow - \infty$. Since $V_{{\rm eff}} \rightarrow 0$ for $r^* \rightarrow \pm \infty$, this implies that the relevant solutions of the wave equation (\ref{schr}) asymptotically scale as
\begin{equation}
\hat{\Psi} \sim e^{-iwr^*}
\label{asy1}
\end{equation}
for $r^* \rightarrow - \infty$ and as
\begin{equation}
\hat{\Psi} \sim A_{{\rm out}} e^{iwr^*} + A_{{\rm in}} e^{-iw r^*}
\label{asy2}
\end{equation}
for $r^* \rightarrow \infty$. Here $A_{{\rm out}}$ and $A_{{\rm in}}$ are complex coefficients that depend on $w$. We want to find the quasinormal mode contribution to the solutions with initial data (\ref{initial}) for large $r$. For this contribution, by definition (\ref{in}), $A_{{\rm in}} = 0$. There are several techniques that can be used to find such solutions. One approach would be to try to use the WKB approximation that we have dealt with in section \ref{section22}, to calculate the reflection and transmission coefficients\footnote{Here, as in scattering of one dimensional quantum mechanics, a reflection coefficient $R$ can be defined as $R \equiv |\frac{A_{\rm in}}{A_{\rm out}}|^2$ and a transmission coefficient $T$ as $T \equiv |\frac{1}{A_{\rm out}}|^2$.} with this method and then to Fourier transform back to real space as done for example in the pioneering work \cite{Vishveshwara:1970zz} (see e.g. \cite{Konoplya:2010kv, Konoplya:2019hlu} for an application of a higher order WKB method along these lines). Another approach which we shall take in this work is to use Greens functions \cite{Detweiler123, Leaver:1986gd, Sun:1988tz, Andersson:1995zk} (see also \cite{Frolov:1998wf, Andersson:2000tf}) and the phase integral method as we will recall in what follows and in appendix \ref{appendixb}. We shall follow the analysis in \cite{Detweiler123, Leaver:1986gd, Sun:1988tz, Andersson:1995zk} and apply it to our case of an axial gravitational wave pulse scattered by a monopole black hole. For completeness we also present most of the steps in the derivation that need no modification when compared to the derivations in \cite{Detweiler123, Leaver:1986gd, Sun:1988tz, Andersson:1995zk}. We mainly adopt the notation from \cite{Andersson:1995zk}.

Since $\Psi$ vanishes for $t < t_0$ because of causality, one can use (\ref{schr}) to define a function $\tilde{\Psi}$ by the integral transform
\begin{equation}
\tilde{\Psi}(r^*, w) \equiv \int_{t_0}^\infty e^{iwt} \Psi(r^*,t)dt \, ,
\end{equation}
which solves the differential equation
\begin{equation}
\left(\partial_{r^*}^2 + w^2 - V_{{\rm eff}}(r)\right)\tilde{\Psi}(r^*, w) = I(r^*, w) \, ,
\label{equationi}
\end{equation}
where
\begin{equation}
I(r^*, w) \equiv e^{iw t_0} \left(iw \Psi(r^*, t_0) - \frac{\partial \Psi(r^*, t)}{\partial t}\bigg|_{t = t_0}\right) = e^{iw t_0} iw ae^{-b\left(r^*-x\right)^2} \, .
\label{i}
\end{equation}
The solution $\tilde{\Psi}(r^*, w)$ can be written in terms of the Greens function $G({r^*}', r^*)$ that solves
\begin{equation}
\left(\partial_{r^*}^2 + w^2 - V_{{\rm eff}}\right)G(r^*, {r^*}') = \delta(r^* - {r^*}')
\end{equation}
as\footnote{The boundary conditions of the Greens function are chosen such that surface terms disappear, see \cite{Andersson:1995zk} for a more detailed discussion about this point.}
\begin{equation}
\tilde{\Psi}(r^*, w) = \int G({r^*}', r^*) I({r^*}', w) d{r^*}' \, .
\label{solj}
\end{equation}
The Greens function can be expressed as linear combination of two independent solutions of the homogeneous equation (\ref{schr}), one, $\hat{\Psi}^-$, that is a purely ingoing wave at the horizon and a linear combination of in- and outgoing waves at infinity and another one, $\hat{\Psi}^+$, that is a purely outgoing wave at infinity and a linear combination of in- and outgoing waves at the horizon. This means
\begin{equation}
\hat{\Psi}^- \sim e^{-iwr^*} \, , \hat{\Psi}^+ \sim B_{{\rm out}} e^{iw r^*} + B_{{\rm in}} e^{-iwr^*} 
\end{equation}
for $r^* \rightarrow -\infty$ and
\begin{equation}
\hat{\Psi}^- \sim A_{{\rm out}} e^{iw r^*} + A_{{\rm in}} e^{-iwr^*} \, , \hat{\Psi}^+ \sim e^{iwr^*}
\end{equation}
for $r^* \rightarrow \infty$.

The solution (\ref{solj}) is then
\begin{equation}
\tilde{\Psi}(r^*, w) = \hat{\Psi}^+ \int_{- \infty}^{r^*} \frac{I \hat{\Psi}^-}{W} d{r^*}' + \hat{\Psi}^- \int_{r^*}^{\infty} \frac{I \hat{\Psi}^+}{W} d{r^*}'  \, ,
\label{so}
\end{equation}
where $W$ is the Wronskian
\begin{equation}
W \equiv \hat{\Psi}^- \partial_{r^*}\hat{\Psi}^+ - \hat{\Psi}^+ \partial_{r^*}\hat{\Psi}^- = 2iw A_{{\rm in}} \, .
\end{equation}
For large $r^*$ (\ref{so}) can be approximated by
\begin{equation}
\tilde{\Psi}(r^*, w) = \frac{J(w)}{2iwA_{{\rm in}}}e^{iw r^*} \, ,
\end{equation}
with
\begin{equation}
J(w) \equiv \int I \hat{\Psi}^- d{r^*}' \, .
\label{jwn}
\end{equation}

Therefore, far away from the black hole the solution to (\ref{equationi}) can be written in the time domain as
\begin{equation}
\Psi(r^*, t) = \frac{1}{4 \pi i}\int_C \frac{e^{-iw\left(t-r^*\right)}}{w A_{{\rm in}}}J(w) dw \, ,
\label{psi}
\end{equation}
where the contour $C$ is given for example in figure 1 in \cite{Andersson:1995zk}. Since we are interested in the quasinormal mode contribution to (\ref{psi}) for which $A_{{\rm in}}(w) = 0$ (\ref{in}), we can approximate $A_{{\rm in}}$ around each quasinormal mode $w_n$ (all values $w_n$ for which $A_{{\rm in}}(w_n) = 0$) as $A_{{\rm in}} \approx (w - w_n) \alpha_n$ where $\alpha_n \equiv \partial_w A_{{\rm in}}(w_n)$. According to the residue theorem, the quasinormal mode contribution is given by\footnote{Note that the contour $C$ encloses all the poles in the complex plane that correspond to the quasinormal modes, see e.g. figure 1 in \cite{Andersson:1995zk}.}
\begin{equation}
\Psi_Q(r^*, t) = - \frac{1}{2} \sum_n \frac{e^{-i w_n (t - r^*)}}{w_n \alpha_n} J(w_n) \, .
\end{equation}
Since quasinormal modes come in pairs $w_n$ and $-w_n^*$ \cite{Kokkotas:1999bd, Konoplya:2011qq} (where the star denotes complex conjugation), this expression can be simplified to
\begin{equation}
\Psi_Q(r^*, t) = - {\rm Re} \left(\sum_n \frac{e^{-i w_n (t - r^*)}}{w_n \alpha_n} J(w_n)\right) \, .
\end{equation}
In the appendix \ref{appendixb} we show that 
\begin{equation}
\alpha_n = - 2 \frac{c}{\sqrt{w_n}} e^{-i \eta(w_n)} \left(\frac{\partial \gamma}{\partial w}\right)\bigg|_{w = w_n}
\label{al}
\end{equation}
and that $J(w_n)$ can for large $r$ be approximated as
\begin{equation}
J(w_n) \approx ic \sqrt{w_n} a e^{i \eta(w_n)} \sqrt{\frac{\pi}{b}} e^{i w_n \left(x+t_0\right) - \frac{w_n^2}{4b}} \, ,
\label{kn}
\end{equation}
where the definitions for the functions $\eta$ and $\gamma$ are given in the appendix \ref{appendixb}. 
Therefore, for large $r$
\begin{equation}
\Psi_Q(r^*, t) \approx \frac{a}{2} \sqrt{\frac{\pi}{b}} {\rm Re}\left(\sum_n ie^{2i \eta(w_n)} e^\frac{-w_n^2}{4b} \left(\frac{\partial \gamma}{\partial w}\right)^{-1}\bigg|_{w = w_n} e^{-i w_n (t-t_0-x-r^*)}\right) \, .
\label{expr}
\end{equation}
Both in the derivation we gave and in the final expression (\ref{expr}) there is no difference when compared to the Schwarzschild case \cite{Andersson:1995zk}. The derivation in the appendix \ref{appendixb} takes care of the fact that we are using a monopole black hole instead of a Schwarzschild black hole. This gives rise to different values for $\eta(w_n)$ and $\partial_w \gamma|_{(w=w_n)}$ when compared to the values in the Schwarzschild case.

\begin{figure}
\includegraphics[scale=0.65]{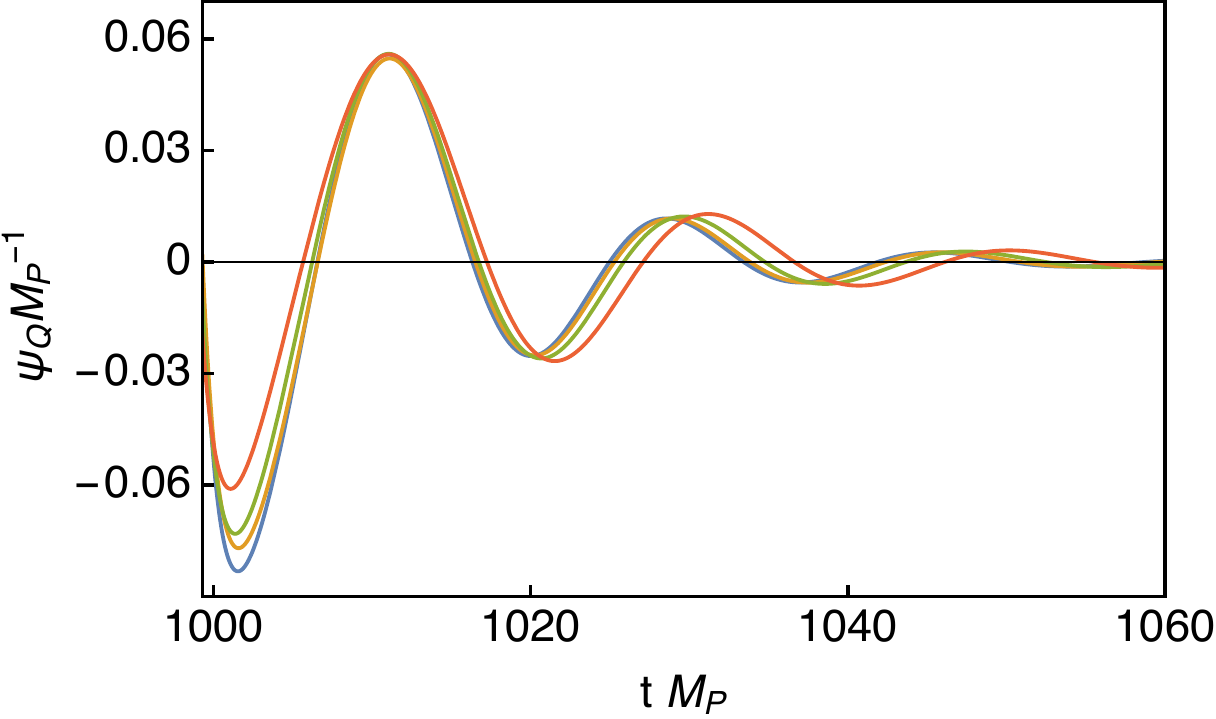}
\includegraphics[scale=0.62]{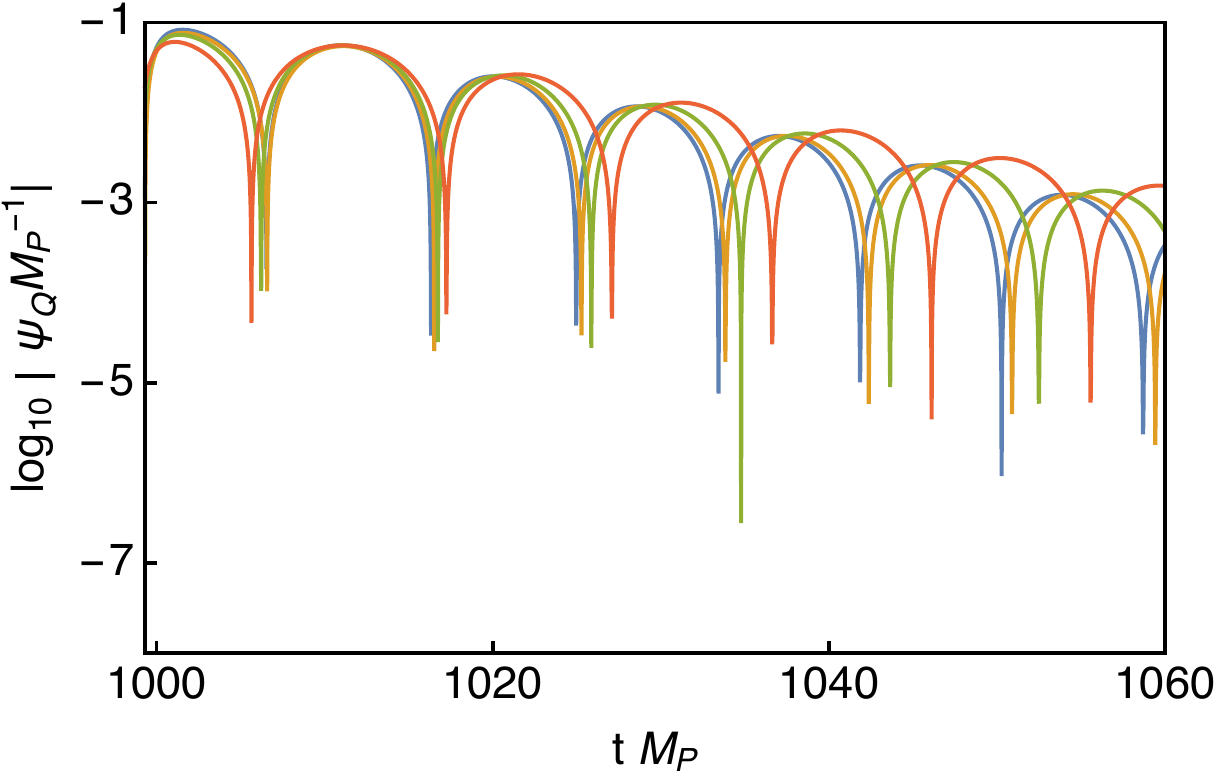}
\caption{The sum of the first four modes ($n=0,1,2,3$) of the quasinormal mode contribution to the response of the monopole black hole to an impinging Gaussian wave pulse for $x+r^*=1000 M_P^{-1}$, $t_0 = 0$, $a=M_P$, $b=M_P^2$, $l=2$ and $M=M_P$. The different colors represent different values of $Nv^2$: The blue line is the Schwarzschild case $Nv^2 = 0$, the yellow line is the case $Nv^2 = 0.01 M_P^2$, the green line is $Nv^2 = 0.04 M_P^2$ and the red line is $Nv^2 = 0.09 M_P^2$.}
\label{figure2}
\end{figure}

\begin{figure}
\includegraphics[scale=0.65]{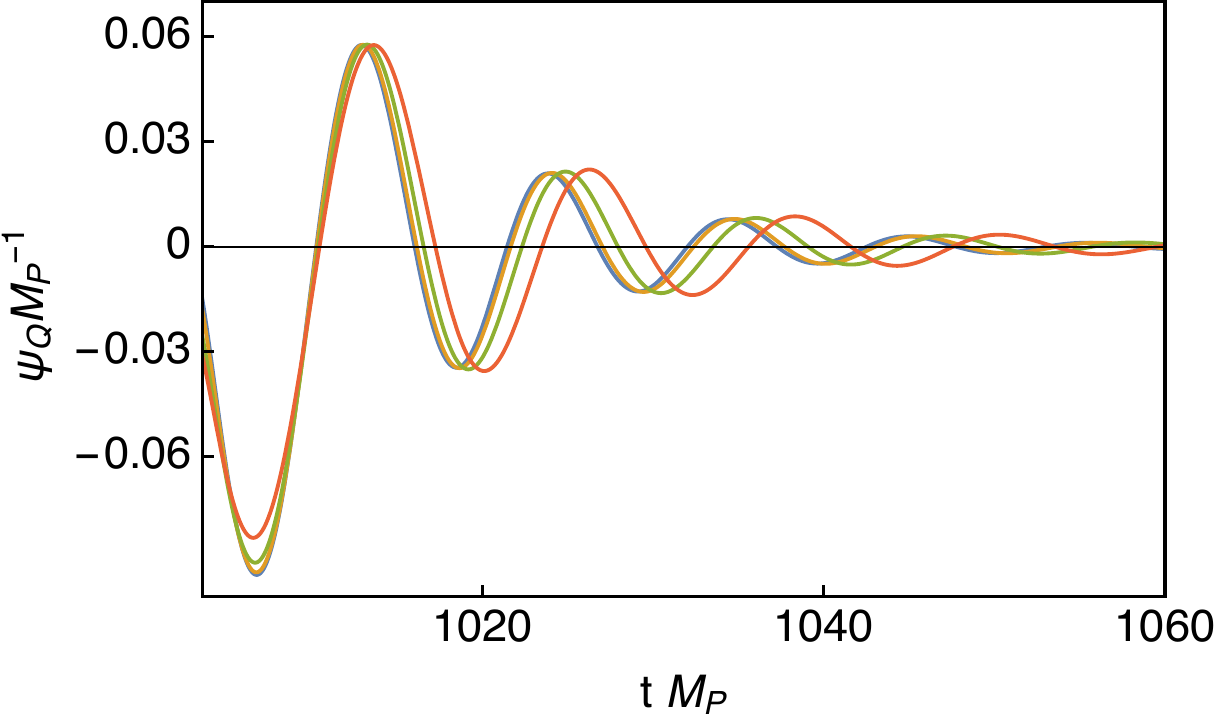}
\includegraphics[scale=0.62]{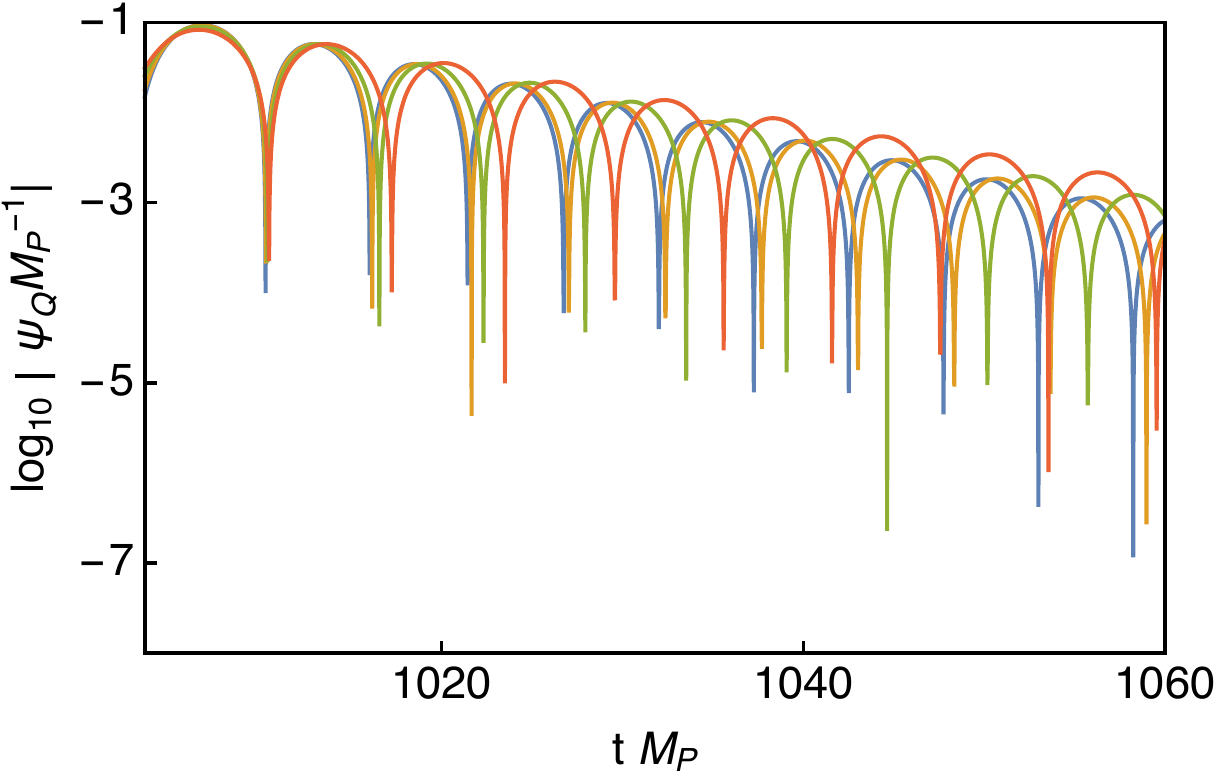}
\caption{The sum of the first four modes ($n=0,1,2,3$) of the quasinormal mode contribution to the response of the monopole black hole to an impinging Gaussian wave pulse for $x+r^*=1000 M_P^{-1}$, $t_0 = 0$, $a = M_P$, $b= M_P^2$, $l=3$ and $M=M_P$. The different colors represent different values of $Nv^2$: The blue line is the Schwarzschild case $Nv^2 = 0$, the yellow line is the case $Nv^2 = 0.01 M_P^2$, the green line is $Nv^2 = 0.04 M_P^2$ and the red line is $Nv^2 = 0.09 M_P^2$.}
\label{figure3}
\end{figure}

We calculated the values of $\eta(w_n)$ and $\partial_w \gamma|_{(w=w_n)}$ for the quasinormal mode frequencies that we have determined in section \ref{section22}. We present the results in table \ref{table1} and plot the corresponding quasinormal mode contribution to the responses of the black holes to impinging pulses with $a = M_P$ and $b=M_P^2$ in figure \ref{figure2} and figure \ref{figure3}. From these figures, one can therefore see how the quasinormal mode contribution to the response of a global monopole black hole to an impinging axial gravitational wave pulse with initial data (\ref{initial}) depends on $Nv^2$. The plots should be seen as results that show the qualitative behavior that we can expect and are not meant to be precise quantitative predictions. This is because we did not obtain these plots from a precise numerical analysis but rather used (semi-)analytic methods in the derivation which required several approximations to be made.

\section{Discussion and outlook} \label{section4}

\subsection{Summary} \label{section41}

We have studied axial tensor quasinormal modes of black holes with $N$ global monopoles inside. We showed how the quasinormal modes depend on $Nv^2$ by calculating the quasinormal mode frequencies for global monopole black holes with several particular values of $Nv^2$. (Here $v$ is the symmetry breaking scale of the model that gave rise to the monopoles.) On top of that, for monopole black holes with these particular values of $Nv^2$, we have studied one relatively simple dynamical process in which these quasinormal modes get excited, the scattering of short pulses of axial gravitational waves by the monopole black holes. We have determined the quasinormal mode contributions to the responses of the monopole black holes to the impinging pulses and in this way showed how the responses depend on $Nv^2$. We used semianalytical methods instead of performing a full numerical analysis. Our results are supposed to show the correct qualitative behavior rather than being precise quantitative predictions. Finally, we have mentioned that the quasinormal modes that we determined (the analogous modes in the case of rotating black holes respectively) are also expected to get excited and to show up in various other dynamical processes such as the ringdown phase of a black hole binary merger in case at least one of the companion black holes is a black hole with global monopoles inside.

\subsection{Observable manifestations} \label{section42}

In regard of the recent and potential future developments in the detection of gravitational waves, it is an interesting question whether or not the effects that we found in our theoretical analysis can have observable consequences, for example in the context of measurements of the ringdown phase of a binary black hole merger with gravitational wave interferometers. Since astrophysical black holes are typically rotating, a necessary step to quantitatively answer this question would be to generalize our analysis that was done for a spherically symmetric monopole black hole to the case of a rotating black hole. We expect that also in that case the black hole's quasinormal modes will depend on on $Nv^2$. When asking for potential observable manifestations in a measurement in which quasinormal modes show up, one should therefore ask what values of $Nv^2$ are realistic for black holes in our universe. The larger $Nv^2$ in our universe can be the bigger potential effects can become. Although it seems difficult to derive or even estimate realistic values of $Nv^2$ without having concrete models and simulations of the formation processes of global monopoles and monopole black holes at hand, one can ask if there are upper bounds on $Nv^2$ that imply constraints on possible observable manifestations. The most naive upper bound is the bound that the maximal possible number $N$ of monopoles that can be inside of a black hole in our universe must be smaller than (or at most equal to) the maximal possible total number $N_H$ of monopoles that can be present in the universe at all.

Let us comment on some (mostly known) points about potential realistic values of $N_H$, $N$ and $v$ in our universe. As we will argue, although some naive constraints can be found, these do not exclude the option that observable effects of global monopole black hole can be present in our universe. A detailed analysis of how large these potential observational effects can be however seems model dependent and remains a task for future work. Another task for future work is to figure out whether there are other (less naive) constraints on $N_H$, $N$ and $v$.

\subsubsection{Formation of global monopoles}

The first question one should ask is, if global monopoles and/or global monopole black holes exist at all in our universe (if $N$, $N_H$ are different from zero). So far, to our knowledge, no signals that imply their existence have been found. As discussed in the course of this work, there exist however simple models in theory space that give rise to global monopoles. These models can potentially be realized in nature and can lead to the formation of global monopoles in our universe. For example, there could be a dark sector with a global symmetry that, when broken at a scale $v$, gives rise to (dark) global monopoles. There could also be a GUT with such a global symmetry in our universe (see e.g. \cite{Davidson:1981zd, Wilczek:1982rv, Gelmini:1982zz} for an example of a GUT with global symmetry)\footnote{Since there are doubts about the existence of fundamental global symmetries in nature (see e.g. \cite{Banks:2010zn} for a relatively recent work), we want to emphasize that the global symmetry that gives rise to such global monopoles can be a global symmetry in a low-energy effective model.}. In that latter case $v$ would be the typical GUT scale. When such a model is realized in nature, global monopoles and antimonopoles could have been formed in a phase transition in the early universe. Usually, it is argued that the monopoles and antimonopoles are produced in pairs \cite{Linde:2005ht} which in particular leads to a finite total energy of the monopoles and antimonopoles (see the discussion in section \ref{section11}). At the formation time they have a typical distance of $H^{-1}(v)$ where $H$ is the value of the Hubble constant at the formation time. Depending on their evolution one can investigate whether or not they can end up in black holes and whether one can constrain their abundance by existing observational data.

\subsubsection{Evolution of global monopoles and formation of monopole black holes}

The evolution history of the global monopoles and antimonopoles depends on the moment of their formation, typically one expects that they are formed at the time of inflation \cite{Linde:2005ht}. During inflation, the monopoles and antimonopoles are then pulled apart. After inflation, when the horizon grows (in the radiation and matter dominated eras), depending on the inflationary model, two different kinds of events that are relevant for our discussion can happen. First, independently of the potential existence of global monopoles, primordial black holes can be formed, for example via their most popular formation mechanism, the collapse of overdense regions once order one fluctuations of primordial origin re-enter the horizon \cite{Zel:1967, Hawking:1971ei, Carr:1974nx}. When pre-existing monopoles happen to be present in those collapsing overdense regions, primordial monopole black holes will be formed. This is the sketch of one formation mechanism of global monopole black holes that seems to be realizable in our universe.\footnote{Such possible kinds of formation mechanism have been discussed in the literature in some detail already for magnetic monopole black holes, see e.g. \cite{Stojkovic:2004hz}. Another potential formation mechanism of monopole black holes might come from the capture of monopoles by existing black holes.} Without a concrete model it is however unclear what fraction of possible primordial black holes can be of the monopole kind as a result of this mechanism. Second, depending on how much the global monopoles and global antimonopoles got diluted by inflation, at some point the monopoles and antimonopoles (and the monopole black holes) can potentially start to re-enter the horizon. After horizon re-entering global monopoles and global antimonopoles will attract each other because of the attractive long range force between them (see the discussion in section \ref{section11}). They approach each other, emit Goldstone bosons and gravitational waves and finally annihilate when they collide. In case they do not interact with matter at all (or at most very weakly), numerical simulations indicate that this process is very efficient: In that case, according to \cite{Bennett:1990xy}, in the radiation dominated era only an average number of $N_H = 3.5 \pm 1.5$ global monopoles per horizon and in the matter dominated era only $N_H = 4.0 \pm 1.5$ global monopoles per horizon will survive.\footnote{See \cite{Yamaguchi:2001rf} for a more recent simulation with slightly different results.} If they however interact with matter strongly, their motion will be damped and many more monopoles and antimonopoles per horizon can survive. It seems natural to expect that such a damping in particular happens in the case which is particularly relevant for us, namely in the case when some of the global monopoles and/or global antimonopoles ended up in primordial black holes before they re-entered the horizon.

\subsubsection{Cosmological and astrophysical bounds}

In case of efficient annihilation, the number $N_H$ remains constant during the radiation dominated era, $N_H = 3.5 \pm 1.5$ (matter dominated era, $N_H = 4.0 \pm 1.5$ resp.) \cite{Bennett:1990xy}. The fluctuations are of order $\sqrt{N_H}$.  The average energy density of global monopoles is therefore given by \cite{Vilenkin:2000jqa}
\begin{equation}
\rho = N_H m d_H^{-3} = N_H \left(4 \pi v^2 n_M^{-\frac{1}{3}}\right) d_H^{-3} \, ,
\end{equation}
where $d_H$ is the Hubble size and $m = 4 \pi v^2 n_M^{-\frac{1}{3}}$ the typical mass of a monopole with $n_M$ the number per horizon volume. Therefore, one factor of $d_H$ cancels out which implies the constant ratio
\begin{equation}
\frac{\delta \rho}{\rho_m} \approx 3 \frac{v^2}{M_P^2} \, ,
\end{equation}
where $\rho_m = 3 H^2 M_P^2$ is the energy density of the matter in our universe. This effect of a constant ratio $\frac{\delta \rho}{\rho_m}$ which is sometimes called ``scaling" \cite{Bennett:1990xy, Pen:1993nx, Yamaguchi:2001rf} is also known from networks of local and global cosmic strings (see e.g. \cite{Kibble:1984hp, Albrecht:1984xv, Bennett:1985qt, Bennett:1986zn, Bennett:1987vf, Albrecht:1989mk, Allen:1990tv, Bennett:1989yp, Martins:1995tg, Martins:1996jp, Vincent:1996rb, Yamaguchi:1998gx, Yamaguchi:1999yp, Yamaguchi:1999dy, Yamaguchi:2002zv}) as well as from global textures (see e.g. \cite{Turok:1989ai, Spergel:1990ee, Turok:1990mi}). In case horizon re-entry of global monopoles and global antimonopoles has happened early enough (before recombination), this has an important consequence: It implies that global monopoles can be sources for scale-invariant density perturbations in the early universe which contribute to the CMB power spectrum \cite{Bennett:1992fy}. In that case, CMB observations can therefore constrain the symmetry breaking scale $v$ of the model that gave rise to the global monopoles \cite{Bennett:1992fy, Durrer:1998rw, Bevis:2004wk, Lopez-Eiguren:2017dmc}. The current bound from PLANCK data is $v <6.4 \times 10^{15} {\rm GeV}$ \cite{Lopez-Eiguren:2017dmc}.

On top of the cosmological considerations, astrophysical bounds on global monopoles exist \cite{Hiscock:1990ev}. One such bound on the existence of global monopoles in our universe exists for the case of relativistic monopoles: Relativistic global monopoles cause huge tidal forces on any nearby orbit \cite{Hiscock:1990ev}. The absence of such tidal forces in our observational data implies that there can be at most one ultrarelativistic global monopole in our local group of galaxies \cite{Hiscock:1990ev}. This bound however seems not to apply for global monopoles that are inside of black holes, particularly when they have been swallowed by the black holes already in the early universe. This is because we do not expect the black holes to be moving with ultrarelativistic speeds. Another astrophysical upper bound on the number of global monopoles in our universe comes from the requirement that global monopoles should not overclose the universe. Because global monopoles have an infinite energy, this bound allows much less global monopoles to be present than ordinary particles. Nevertheless, this requirement still allows many global monopoles to be present within the current Hubble horizon \cite{Hiscock:1990ev}.

\subsubsection{Conclusion}

In conclusion, although we can get some ideas about possible values of $v$ and $N_H$ in our universe from cosmological and astrophysical considerations (which can imply very naive upper bounds on possible values of $Nv^2$ for monopole black holes), such considerations seem model dependent: One important ingredient in the above-recalled cosmological arguments is the moment of horizon re-entry of global monopoles and global antimonopoles. This moment depends on the formation time of global monopoles, on the inflationary model and mainly on the duration of the part of inflation after monopole formation time. Another important ingredient is the process of monopole and antimonopole annihilation. Although this annihilation process is very efficient in case global monopoles interact with matter very weakly (with the result that only order one global monopoles and global antimonopoles per horizon survive \cite{Bennett:1990xy}), it is unclear how much the annihilation process is damped when global monopoles and global antimonopoles interact strongly with matter, for example because they happened to be inside of primordial black holes. How many monopoles and antimonopoles ended up in primordial black holes depends on the mechanism of primordial black hole formation and in particular on the number density of global monopoles at the formation time of primordial black holes.

Therefore, both from the cosmological and from the astrophysical considerations that we have discussed, it seems that black holes with a large number of global monopoles inside cannot be excluded a priori in a model independent way. This implies that, given these considerations, one cannot exclude the option that monopole black holes with a number $Nv^2$ that leads to observable effects in processes in which the black hole's quasinormal modes get excited exist in our universe. It is an interesting question of model building if realistic models which are not too fine-tuned can predict such effects and how large the potential effects can become. It would also be interesting to investigate possible other constraints that we have not discussed and that could potentially lead to bounds on $Nv^2$ in our universe.

\subsection{Application to different kinds of non-vacuum black holes} \label{section43}

As discussed in section \ref{section22}, we have used (\ref{metric}), (\ref{Ar}), (\ref{emom}) in our analysis which provide a good approximation to large global monopole black holes. In other regimes detailed numerical solutions \cite{Liebling:1999bb, Maison:1999ke, Nucamendi:2000af, Tamaki:2003kv} have to be considered. These solutions have a significantly different near-horizon geometry when compared to (\ref{metric}). It is an interesting question for future work if and how our analysis can be applied to those numerical solutions and in particular how the quasinormal modes change as a consequence of the different near-horizon geometry.

Analogous numerical black hole solutions with modified near-horizon geometry also exist for different kinds of matter in certain parameter domains (see e.g. \cite{Herdeiro:2015waa} for a recent review), for example for local magnetic monopoles (see e.g. \cite{Ortiz:1991eu, Lee:1991vy, Breitenlohner:1991aa, Breitenlohner:1994di, Aichelburg:1992st, Lee:1994sk, Breitenlohner:1997hm, Galtsov:1997lak, Breitenlohner:1997ud, Hartmann:2001ic, Brihaye:2002km, Kleihaus:2004gm, Maldacena:2020skw, Ghosh:2020tdu, Bai:2020spd}) instead of global monopoles as studied here. Whereas scattering of plane (scalar) waves by several kinds of those non-vacuum black holes has already been studied \cite{Dvali:2016mur, Gussmann:2016mkp}, it is an interesting question for future work how, in those cases, scattering processes of short pulses of gravitational waves and the corresponding tensor quasinormal modes get modified as a consequence of the structure of the near horizon geometries. Such modifications can potentially have important consequences in particular for the context of the ringdown phase of a binary black hole merger in case at least one of the companion black holes is a non-vacuum black hole of that kind.

\section*{Acknowledgement}

We thank Artem Averin for some discussions. We thank Roman Konoplya for comments, in particular for pointing out reference \cite{Konoplya:2019hlu} and for making the Mathematica package \cite{code} publicly available to download. This work was supported by the Deutsche Forschungsgemeinschaft (DFG, German Research Foundation) under the individual grant GU 2052/1-1.

\appendix

\section{Components of the axial perturbations of the Einstein tensor} \label{appendixa}

In this appendix we provide the expressions for the components of the axial perturbations of the Einstein tensor, $\delta G_{\mu \nu}^{{\rm (axial)}}$, that we have used when deriving (\ref{eq1}), (\ref{eq2}) and (\ref{eq3}). The background Einstein field equations are used when deriving these components. The functions $\hat{h}_0$ and $\hat{h}_1$ depend only on the radial coordinate $r$ and on the frequency $w$.
\begin{align}
\begin{split}
\delta G_{\theta \phi}^{{\rm (axial)}} =& \frac{1}{4} \left(-2iA(r)^{-1} w \hat{h}_0 - \frac{4 M}{M_P^2 r^2} \hat{h}_1 - 2A(r) \partial_r \hat{h}_1\right) \\
& e^{-iwt} {\rm sin} \theta \left(\partial_\theta^2 - \rm{cot} \theta \partial_\theta - \frac{1}{{\rm sin} \theta} \partial_\phi^2\right)\left(P_l({\rm cos} \theta) e^{im \phi}\right)
\end{split}
\end{align}
\begin{align}
\begin{split}
\delta G_{r \phi}^{{\rm (axial)}} =& \bigg(\frac{1}{2r}A(r)^{-1}\left(2iw \hat{h}_0+rw^2 \hat{h}_1 - iwr \partial_r \hat{h}_0 \right) \\
&- \frac{1}{2r^2}l(l+1) \hat{h}_1 +\frac{1}{r^2} \hat{h}_1 \bigg) {\rm sin} \theta \partial_\theta P_l({\rm cos} \theta) e^{-iwt + im\phi}
\end{split}
\end{align}
\begin{align}
\begin{split}
\delta G_{t \phi}^{{\rm (axial)}} =& \bigg(\frac{1}{2}A(r) \partial_r^2 \hat{h}_0 + \frac{1}{2} iw A(r) \partial_r \hat{h}_1 +\frac{iw}{r}A(r) \hat{h}_1 - \frac{1}{2r^2}l(l+1)\hat{h}_0 \\
&+ \frac{2M}{M_P^2 r^3} \hat{h}_0 + \frac{N}{M_P^2} \frac{v^2}{r^2}\hat{h}_0\bigg) {\rm sin} \theta \partial_\theta P_l({\rm cos} \theta) e^{-iwt + im\phi}
\end{split}
\end{align}

\section{Derivation of (\ref{al}) and (\ref{kn})} \label{appendixb}

In this appendix we derive (\ref{al}) and (\ref{kn}) and give the expressions for $\eta$ and $\gamma$ that are used in (\ref{expr}). We shall use the phase integral method \cite{Andersson:1995zk, Andersson:19, Andersson:1992scr, Froeman:1992gp} that was used in \cite{Andersson:1995zk} to determine the quasinormal mode contribution to the response of a Schwarzschild black hole to a Gaussian scalar wave pulse. We follow the derivation in \cite{Andersson:1995zk} and apply it to the case of axial gravitational waves and global monopole black holes. We mainly adopt the notation from \cite{Andersson:1995zk}.

As a first step, it is convenient to write the wave equation (\ref{schr}) that we want to solve in terms of the radial coordinate $r$ instead of the tortoise coordinate $r^*$. This gives
\begin{equation}
\left(\partial_r^2 + R(r)\right)\hat{\Phi} = 0 \, ,
\label{eq}
\end{equation}
where
\begin{equation}
\hat{\Phi} \equiv \sqrt{A(r)} \hat{\Psi} \, ,
\end{equation}
\begin{equation}
R(r) \equiv \frac{1}{A(r)^2}\left(w^2 - V_{{\rm eff}} +\frac{M^2}{M_P^4 r^4} + \frac{2M}{M_P^2 r^3}A(r)\right) \, .
\end{equation}

Since we demand the absence of outgoing waves at the event horizon, the relevant solutions to (\ref{eq}) asymptotically scale as

\begin{equation}
\hat{\Phi} \sim \left(\frac{r M_P^2 \left(1-\frac{N v^2}{M_P^2}\right)}{2M}-1\right)^{\frac{1}{2} - \frac{2iMw}{M_P^2\left(1-\frac{Nv^2}{M_P^2}\right)^2}} e^{\frac{-iwr}{1-\frac{Nv^2}{M_P^2}}}
\end{equation}
for $r \rightarrow \frac{2M}{M_P^2 \left(1-\frac{Nv^2}{M_P^2}\right)}$ (see (\ref{asy1})) and as
\begin{align}
\begin{split}
\hat{\Phi} \sim& A_{{\rm out}} \sqrt{\left(1-\frac{Nv^2}{M_P^2}\right)} e^{\frac{iwr}{1-\frac{Nv^2}{M_P^2}}} \left(\frac{rM_P^2 \left(1-\frac{Nv^2}{M_P^2}\right)}{2M}\right)^{\frac{2iwM}{M_P^2\left(1-\frac{Nv^2}{M_P^2}\right)^2}} \\
&+ A_{{\rm in}} \sqrt{\left(1-\frac{Nv^2}{M_P^2}\right)} e^{\frac{-iwr}{1-\frac{Nv^2}{M_P^2}}}\left(\frac{rM_P^2\left(1-\frac{Nv^2}{M_P^2}\right)}{2M}\right)^{\frac{-2iwM}{M_P^2\left(1-\frac{Nv^2}{M_P^2}\right)^2}}
\label{asyminf}
\end{split}
\end{align}
for $r \rightarrow \infty$ (see (\ref{asy2})).

As a next step one can find an approximate solution to (\ref{eq}) by using the phase-integral method \cite{Andersson:1995zk, Andersson:19, Andersson:1992scr, Froeman:1992gp}. To lowest order approximation, the general solution of (\ref{eq}) can be written as a linear combinations of the two functions $f_1$ and $f_2$ which are given by \cite{Andersson:1995zk}
\begin{equation}
f_{1, 2}(r, t_j) = \frac{1}{\sqrt{Q(r)}} e^{\pm i \int_{t_j}^r Q(r')} dr' \, .
\label{phase}
\end{equation}
Here the sign ``$+$" corresponds to $f_1$ whereas ``$-$" corresponds to $f_2$. $t_j$ are the zeros of the function $Q^2(r)$. $Q^2(r)$ is given by
\begin{equation}
Q^2(r) \equiv R(r) - \frac{1}{4 \left(r-\frac{2M}{M_P^2\left(1-\frac{Nv^2}{M_P^2}\right)}\right)^2} \, .
\label{term}
\end{equation}
The second term is usually added in these contexts for the approximation to match the scaling-behavior of the exact solution in the near-horizon regime \cite{Andersson:19}. This is because $R(r)$ has a second order pole at $r = \frac{2M}{M_P^2\left(1-\frac{Nv^2}{M_P^2}\right)}$ which implies that the exact solution of (\ref{eq}) scales for $r \rightarrow \frac{2M}{M_P^2 \left(1-\frac{Nv^2}{M_P^2}\right)}$ as \cite{Andersson:19}
\begin{equation}
\hat{\Phi} \sim \left(r-\frac{2M}{M_P^2\left(1-\frac{Nv^2}{M_P^2}\right)}\right)^{\alpha+1}, \hat{\Phi} \sim \left(r-\frac{2M}{M_P^2\left(1-\frac{Nv^2}{M_P^2}\right)}\right)^{-\alpha} \, ,
\end{equation}
where $\alpha$ is defined by
\begin{equation}
\alpha \left(\alpha + 1\right) \equiv -\left(\frac{2M}{M_P^2\left(1-\frac{Nv^2}{M_P^2}\right)}\right)^2 \frac{1}{\left(1-\frac{Nv^2}{M_P^2}\right)^2}\left(w^2 + \frac{M^2M_P^2\left(1-\frac{Nv^2}{M_P^2}\right)^4}{(2M)^4}\right) \, .
\end{equation}
It is easy to see that the phase integral approximation (\ref{phase}) matches this behavior when the second term in (\ref{term}) is added.

Using (\ref{term}), for large $r$ the functions $f_{1,2}(r, t_1)$ can be approximated as
\begin{equation}
f_{1,2}(r,t_1) \approx \frac{\sqrt{1-\frac{Nv^2}{M_P^2}}}{\sqrt{w}} \left(\frac{rM_P^2\left(1-\frac{Nv^2}{M_P^2}\right)}{2M}\right)^{\pm \frac{2iMw}{M_P^2 \left(1-\frac{Nv^2}{M_P^2}\right)^2}}e^{\pm i\left(\frac{w}{1-\frac{Nv^2}{M_P^2}}r + \eta(w)\right)} \, ,
\label{asy}
\end{equation}
where
\begin{align}
\begin{split}
\eta(w) \equiv& \int_{t_1}^\infty \left(Q(r) - w \left(1-\frac{2M}{M_P^2 r}-\frac{Nv^2}{M_P^2}\right)^{-1}\right) dr \\
&- \frac{w}{1-\frac{Nv^2}{M_P^2}}\left(t_1 + \frac{2M}{M_P^2\left(1-\frac{Nv^2}{M_P^2}\right)} {\rm ln} \left(\frac{t_1M_P^2\left(1-\frac{Nv^2}{M_P^2}\right)}{2M}-1\right)\right) \, .
\end{split}
\end{align}
We evaluated this integral to calculate $\eta$ for the quasinormal mode frequencies $w_n$ that we have determined in section \ref{section22}. The results for $\eta(w_n)$ are listed in table \ref{table1}.

As discussed in \cite{Andersson:1995zk}, for large $r$ the linear approximation of $f_1$ and $f_2$ that solves (\ref{eq}) is 
\begin{equation}
\hat{\Phi} \approx c \left(-ie^{i\gamma}f_1(r,t_1) + \left(e^{i\gamma} + e^{-i \gamma}\right)f_2(r, t_1)\right) \, ,
\end{equation}
where $c$ is a normalisation constant and 
\begin{equation}
\gamma \equiv \int_{t_2}^{t_1} Q dr \, .
\end{equation}

Inserting (\ref{asy}) gives for large $r$
\begin{align}
\begin{split}
\hat{\Phi} \approx& \frac{c \sqrt{1-\frac{Nv^2}{M_P^2}}}{\sqrt{w}} \Bigg(-ie^{i\left(\gamma + \eta\right)} \left(\frac{rM_P^2\left(1-\frac{Nv^2}{M_P^2}\right)}{2M}\right)^{\frac{2iMw}{M_P^2\left(1-\frac{Nv^2}{M_P^2}\right)^2}} e^{\frac{iwr}{1-\frac{Nv^2}{M_P^2}}}\\
&+ \left(e^{i \gamma} +e^{-i \gamma}\right) e^{-i \eta} \left(\frac{rM_P^2\left(1-\frac{Nv^2}{M_P^2}\right)}{2M}\right)^{-\frac{2iMw}{M_P^2\left(1-\frac{Nv^2}{M_P^2}\right)^2}}e^{-\frac{iwr}{1-\frac{Nv^2}{M_P^2}}}\Bigg) \, .
\end{split}
\label{asymptoticform}
\end{align}

When this expression is compared to (\ref{asyminf}), one obtains
\begin{equation}
A_{{\rm out}} = \frac{ic}{\sqrt{w}} e^{i\left(\gamma + \eta\right)} \, ,
\end{equation}
\begin{equation}
A_{{\rm in}} = \frac{c}{\sqrt{w}} e^{-i \eta}\left(e^{i \gamma} + e^{-i \gamma}\right) \, .
\label{ain}
\end{equation}
Since $A_{{\rm in}} = 0$ for quasinormal modes, $e^{-2i\gamma(w_n)} = -1$. Thus, for quasinormal mode frequencies $w_n$\footnote{Using (\ref{sommerfeld}) gives rise to another way to determine the quasinormal mode frequencies $w_n$ that we have determined in section \ref{section22} \cite{Andersson:1995zk, Andersson:19, Andersson:1992scr, Froeman:1992gp}.}
\begin{equation}
\gamma (w_n) = \left(n + \frac{1}{2}\right) \pi \, .
\label{sommerfeld}
\end{equation}
By iterating this equation we have determined the values of $\frac{\partial \gamma}{\partial w} \big|_{w = w_n}$ for the quasinormal mode frequencies that we have calculated in section (\ref{section22}). The results are listed in table \ref{table1}.

Taking the derivative of (\ref{ain}) gives for $\alpha_n \equiv \partial_w A_{{\rm in}}(w_n)$ the expression
\begin{equation}
\alpha_n = 2 i \frac{c}{\sqrt{w_n}} e^{i\left(\gamma(w_n) - \eta(w_n)\right)} \left(\frac{\partial \gamma}{\partial w}\right)\bigg|_{w = w_n} \, ,
\end{equation}
which, using (\ref{sommerfeld}), gives (\ref{al}).

Finally, the function $J(w_n)$ that was defined in (\ref{jwn}) with $I$ as in (\ref{i}) can be approximated at large $r$ by using the asymptotic form (\ref{asymptoticform}) with quasinormal mode frequencies $w_n$ inserted. Taking into account (\ref{sommerfeld}), the result is
\begin{equation}
J(w_n) \approx i c e^{i w_n t_0} e^{i \eta(w_n)} \int dr^* \left(1-\frac{Nv^2}{M_P^2}\right)^{- \frac{1}{2}} a e^{-b \left(r^*-x\right)^2} e^{iw_n r^*} \sqrt{w_n\left(1-\frac{Nv^2}{M_P^2}\right)} \, ,
\end{equation}
which yields the expression (\ref{kn}).


\begin{thebibliography}{9}

\bibitem{Vishveshwara:1970zz}
C.~V.~Vishveshwara,
``Scattering of Gravitational Radiation by a Schwarzschild Black-hole,''
Nature \textbf{227}, 936-938 (1970)
doi:10.1038/227936a0

\bibitem{Vishveshwara:1996jgz}
C.~V.~Vishveshwara,
``On the black hole trail ...,''
Proceedings, 18th Conference of the Indian Association for General Relativity and Gravitation (1996)

\bibitem{Andersson:1995zk}
N.~Andersson,
``Excitation of Schwarzschild black hole quasinormal modes,''
Phys. Rev. D \textbf{51}, 353-363 (1995)
doi:10.1103/PhysRevD.51.353

\bibitem{Press:1971wr}
W.~H.~Press,
``Long Wave Trains of Gravitational Waves from a Vibrating Black Hole,''
Astrophys. J. Lett. \textbf{170}, L105-L108 (1971)
doi:10.1086/180849

\bibitem{Kokkotas:1999bd}
K.~D.~Kokkotas and B.~G.~Schmidt,
``Quasinormal modes of stars and black holes,''
Living Rev. Rel. \textbf{2}, 2 (1999)
doi:10.12942/lrr-1999-2
[arXiv:gr-qc/9909058 [gr-qc]].

\bibitem{Berti:2009kk}
E.~Berti, V.~Cardoso and A.~O.~Starinets,
``Quasinormal modes of black holes and black branes,''
Class. Quant. Grav. \textbf{26}, 163001 (2009)
doi:10.1088/0264-9381/26/16/163001
[arXiv:0905.2975 [gr-qc]].

\bibitem{Konoplya:2011qq}
R.~A.~Konoplya and A.~Zhidenko,
``Quasinormal modes of black holes: From astrophysics to string theory,''
Rev. Mod. Phys. \textbf{83}, 793-836 (2011)
doi:10.1103/RevModPhys.83.793
[arXiv:1102.4014 [gr-qc]].

\bibitem{Andersson:1996pn}
N.~Andersson and K.~D.~Kokkotas,
``Gravitational waves and pulsating stars: What can we learn from future observations?,''
Phys. Rev. Lett. \textbf{77}, 4134-4137 (1996)
doi:10.1103/PhysRevLett.77.4134
[arXiv:gr-qc/9610035 [gr-qc]].

\bibitem{Allen:1997xj}
G.~Allen, N.~Andersson, K.~D.~Kokkotas and B.~F.~Schutz,
``Gravitational waves from pulsating stars: Evolving the perturbation equations for a relativistic star,''
Phys. Rev. D \textbf{58}, 124012 (1998)
doi:10.1103/PhysRevD.58.124012
[arXiv:gr-qc/9704023 [gr-qc]].

\bibitem{Hai:2013ara}
H.~Huang, Y.~J.~Wang and J.~H.~Chen,
``Absorption cross section of black holes with global monopole,''
Chin. Phys. B \textbf{22}, no.7, 070401 (2013)
doi:10.1088/1674-1056/22/7/070401

\bibitem{Anacleto:2017kmg}
M.~A.~Anacleto, F.~A.~Brito, S.~J.~S.~Ferreira and E.~Passos,
``Absorption and scattering of a black hole with a global monopole in f(R) gravity,''
Phys. Lett. B \textbf{788}, 231-237 (2019)
doi:10.1016/j.physletb.2018.11.020
[arXiv:1701.08147 [hep-th]].

\bibitem{Pitelli:2017bgx}
J.~P.~M.~Pitelli, V.~S.~Barroso and M.~Richartz,
``Scattering Cross Section and Stability of Global Monopoles,''
Phys. Rev. D \textbf{96}, no.10, 105021 (2017)
doi:10.1103/PhysRevD.96.105021
[arXiv:1711.03526 [gr-qc]].

\bibitem{Graca:2015jea}
J.~P.~Morais Graça, H.~S.~Vieira and V.~B.~Bezerra,
``Scalar and spinor QNMs of a black hole with a global monopole in f(R) gravity,''
Gen. Rel. Grav. \textbf{48}, no.4, 38 (2016)
doi:10.1007/s10714-016-2024-7
[arXiv:1510.07184 [gr-qc]].

\bibitem{Xi:2008ce}
P.~Xi,
``Quasinormal modes of a black hole with quintessence-like matter and a deficit solid angle,''
Astrophys. Space Sci. \textbf{321}, 47-51 (2009)
doi:10.1007/s10509-009-9994-9
[arXiv:0810.4022 [gr-qc]].

\bibitem{Pol:1974}
A.~M.~Polyakov,
``Particle spectrum in quantum field theory,"
JETP. Lett. \textbf{20}, 194 (1974)

\bibitem{Barriola:1989hx}
M.~Barriola and A.~Vilenkin,
``Gravitational Field of a Global Monopole,''
Phys. Rev. Lett. \textbf{63}, 341 (1989)
doi:10.1103/PhysRevLett.63.341

\bibitem{tHooft:1974kcl}
G.~'t Hooft,
``Magnetic Monopoles in Unified Gauge Theories,''
Nucl. Phys. B \textbf{79}, 276-284 (1974)
doi:10.1016/0550-3213(74)90486-6

\bibitem{Goldhaber:1989na}
A.~S.~Goldhaber,
``Collapse of a 'Global Monopole.',''
Phys. Rev. Lett. \textbf{63}, 2158 (1989)
doi:10.1103/PhysRevLett.63.2158

\bibitem{Rhie:1990kc}
S.~H.~Rhie and D.~P.~Bennett,
``Global monopoles do not 'collapse',''
Phys. Rev. Lett. \textbf{67}, 1173 (1991)
doi:10.1103/PhysRevLett.67.1173

\bibitem{Perivolaropoulos:1991du}
L.~Perivolaropoulos,
``Instabilities and interactions of global topological defects,''
Nucl. Phys. B \textbf{375}, 665-693 (1992)
doi:10.1016/0550-3213(92)90115-R

\bibitem{Achucarro:2000td}
A.~Achucarro and J.~Urrestilla,
``The (In)stability of global monopoles revisited,''
Phys. Rev. Lett. \textbf{85}, 3091-3094 (2000)
doi:10.1103/PhysRevLett.85.3091
[arXiv:hep-ph/0003145 [hep-ph]].

\bibitem{Bennett:1990xy}
D.~P.~Bennett and S.~H.~Rhie,
``Cosmological evolution of global monopoles and the origin of large scale structure,''
Phys. Rev. Lett. \textbf{65}, 1709-1712 (1990)
doi:10.1103/PhysRevLett.65.1709

\bibitem{Harari:1990cz}
D.~Harari and C.~Lousto,
``Repulsive gravitational effects of global monopoles,''
Phys. Rev. D \textbf{42}, 2626-2631 (1990)
doi:10.1103/PhysRevD.42.2626

\bibitem{Nucamendi:2000is}
U.~Nucamendi, M.~Salgado and D.~Sudarsky,
``Nonminimal global monopoles and bound orbits,''
Phys. Rev. Lett. \textbf{84}, 3037 (2000)
doi:10.1103/PhysRevLett.84.3037
[arXiv:gr-qc/0002001 [gr-qc]].

\bibitem{Nucamendi:2000ch}
U.~Nucamendi, M.~Salgado and D.~Sudarsky,
``Global monopoles non-minimally coupled to gravity and astrophysical implications,''
AIP Conf. Proc. \textbf{555}, no.1, 363 (2001)
doi:10.1063/1.1363543

\bibitem{Dvali:2000ty}
G.~R.~Dvali,
``Cosmological constant and Fermi-Bose degeneracy,''
[arXiv:hep-th/0004057 [hep-th]].

\bibitem{Spinelly:2002mt}
J.~Spinelly, E.~R.~Bezerra de Mello and U.~De Freitas,
``Gravitating magnetic monopole in the global monopole space-time,''
Phys. Rev. D \textbf{66}, 024018 (2002)
doi:10.1103/PhysRevD.66.024018
[arXiv:hep-th/0205046 [hep-th]].

\bibitem{Achucarro:2002gg}
A.~Achucarro and J.~Urrestilla,
``Comment on `Gravitating magnetic monopole in the global monopole space-time',''
Phys. Rev. D \textbf{68}, 088701 (2003)
doi:10.1103/PhysRevD.68.088701
[arXiv:hep-th/0212148 [hep-th]].

\bibitem{BezerradeMello:2003yt}
E.~R.~Bezerra de Mello,
``Reply on comment on `Gravitating magnetic monopole in the global monopole space-time',''
Phys. Rev. D \textbf{68}, 088702 (2003)
doi:10.1103/PhysRevD.68.088702
[arXiv:hep-th/0304029 [hep-th]].

\bibitem{Liebling:1999bb}
S.~L.~Liebling,
``Static gravitational global monopoles,''
Phys. Rev. D \textbf{61}, 024030 (2000)
doi:10.1103/PhysRevD.61.024030
[arXiv:gr-qc/9906014 [gr-qc]].

\bibitem{Maison:1999ke}
D.~Maison,
``Gravitational global monopoles with horizons,''
[arXiv:gr-qc/9912100 [gr-qc]].

\bibitem{Nucamendi:2000af}
U.~Nucamendi and D.~Sudarsky,
``Black holes with zero mass,''
Class. Quant. Grav. \textbf{17}, 4051-4058 (2000)
doi:10.1088/0264-9381/17/19/306
[arXiv:gr-qc/0004068 [gr-qc]].

\bibitem{Tamaki:2003kv}
T.~Tamaki and N.~Sakai,
``Properties of global monopoles with an event horizon,''
Phys. Rev. D \textbf{69}, 044018 (2004)
doi:10.1103/PhysRevD.69.044018
[arXiv:gr-qc/0309068 [gr-qc]].

\bibitem{Watabe:2004tq}
H.~Watabe and T.~Torii,
``Perturbations of global monopoles as a black hole's hair,''
JCAP \textbf{02}, 001 (2004)
doi:10.1088/1475-7516/2004/02/001
[arXiv:gr-qc/0307074 [gr-qc]].

\bibitem{Regge:1957td}
T.~Regge and J.~A.~Wheeler,
``Stability of a Schwarzschild singularity,''
Phys. Rev. \textbf{108}, 1063-1069 (1957)
doi:10.1103/PhysRev.108.1063

\bibitem{Chandrasekhar:1991fi}
S.~Chandrasekhar and V.~Ferrari,
``On the non-radial oscillations of a star,''
Proc. Roy. Soc. Lond. A \textbf{432}, 247-279 (1991)
doi:10.1098/rspa.1991.0016

\bibitem{Schutz:1985km}
B.~F.~Schutz and C.~M.~Will,
``Black hole normal modes: A semianalytic approach,''
Astrophys. J. Lett. \textbf{291}, L33-L36 (1985)
doi:10.1086/184453

\bibitem{Konoplya:2019hlu}
R.~A.~Konoplya, A.~Zhidenko and A.~F.~Zinhailo,
``Higher order WKB formula for quasinormal modes and grey-body factors: recipes for quick and accurate calculations,''
Class. Quant. Grav. \textbf{36}, 155002 (2019)
doi:10.1088/1361-6382/ab2e25
[arXiv:1904.10333 [gr-qc]].

\bibitem{Iyer:1986np}
S.~Iyer and C.~M.~Will,
``Black Hole Normal Modes: A {WKB} Approach. 1. Foundations and Application of a Higher Order {WKB} Analysis of Potential Barrier Scattering,''
Phys. Rev. D \textbf{35}, 3621 (1987)
doi:10.1103/PhysRevD.35.3621

\bibitem{Matyjasek:2017psv}
J.~Matyjasek and M.~Opala,
``Quasinormal modes of black holes. The improved semianalytic approach,''
Phys. Rev. D \textbf{96}, no.2, 024011 (2017)
doi:10.1103/PhysRevD.96.024011
[arXiv:1704.00361 [gr-qc]].

\bibitem{Matyjasek:2019eeu}
J.~Matyjasek and M.~Telecka,
``Quasinormal modes of black holes. II. Padé summation of the higher-order WKB terms,''
Phys. Rev. D \textbf{100}, no.12, 124006 (2019)
doi:10.1103/PhysRevD.100.124006
[arXiv:1908.09389 [gr-qc]].

\bibitem{Hatsuda:2019eoj}
Y.~Hatsuda,
``Quasinormal modes of black holes and Borel summation,''
Phys. Rev. D \textbf{101}, no.2, 024008 (2020)
doi:10.1103/PhysRevD.101.024008
[arXiv:1906.07232 [gr-qc]].

\bibitem{code}
Mathematica package,
https://goo.gl/nykYGL

\bibitem{Konoplya:2003ii}
R.~A.~Konoplya,
``Quasinormal behavior of the d-dimensional Schwarzschild black hole and higher order WKB approach,''
Phys. Rev. D \textbf{68}, 024018 (2003)
doi:10.1103/PhysRevD.68.024018
[arXiv:gr-qc/0303052 [gr-qc]].

\bibitem{Leaver:1985ax}
E.~W.~Leaver
``An Analytic representation for the quasi normal modes of Kerr black holes,''
Proc. Roy. Soc. Lond. A \textbf{402}, 285-298 (1985)
doi:10.1098/rspa.1985.0119

\bibitem{Andersson947}
N.~Andersson,
``A Numerically accurate investigation of black-hole normal modes,''
Proc. Roy. Soc. Lond. A \textbf{439}, 47-58 (1992)
doi:10.1098/rspa.1992.0133

\bibitem{Kokkotas:1988fm}
K.~D.~Kokkotas and B.~F.~Schutz,
``Black Hole Normal Modes: A {WKB} Approach. 3. The {Reissner-Nordstrom} Black Hole,''
Phys. Rev. D \textbf{37}, 3378-3387 (1988)
doi:10.1103/PhysRevD.37.3378

\bibitem{Andersson:1996cm}
N.~Andersson,
``Evolving test fields in a black hole geometry,''
Phys. Rev. D \textbf{55}, 468-479 (1997)
doi:10.1103/PhysRevD.55.468
[arXiv:gr-qc/9607064 [gr-qc]].

\bibitem{Frolov:1998wf}
V.~P.~Frolov and I.~D.~Novikov,
``Black hole physics: Basic concepts and new developments, Chapter 4''
Fundam. Theor. Phys. \textbf{96} (1998)
doi:10.1007/978-94-011-5139-9

\bibitem{Andersson:2000tf}
N.~Andersson and B.~P.~Jensen,
``Scattering by black holes. Chapter 0.1,''
[arXiv:gr-qc/0011025 [gr-qc]].

\bibitem{Leaver:1986gd}
E.~W.~Leaver,
``Spectral decomposition of the perturbation response of the Schwarzschild geometry,''
Phys. Rev. D \textbf{34}, 384-408 (1986)
doi:10.1103/PhysRevD.34.384

\bibitem{Price:1971fb}
R.~H.~Price,
``Nonspherical perturbations of relativistic gravitational collapse. 1. Scalar and gravitational perturbations,''
Phys. Rev. D \textbf{5}, 2419-2438 (1972)
doi:10.1103/PhysRevD.5.2419

\bibitem{Price:1972pw}
R.~H.~Price,
``Nonspherical Perturbations of Relativistic Gravitational Collapse. II. Integer-Spin, Zero-Rest-Mass Fields,''
Phys. Rev. D \textbf{5}, 2439-2454 (1972)
doi:10.1103/PhysRevD.5.2439

\bibitem{Cunningham:1978zfa}
C.~T.~Cunningham, R.~H.~Price and V.~Moncrief,
``Radiation from collapsing relativistic stars. I - Linearized odd-parity radiation,''
Astrophys. J. \textbf{224}, 643 (1978)
doi:10.1086/156413

\bibitem{Konoplya:2010kv}
R.~A.~Konoplya and A.~Zhidenko,
``Passage of radiation through wormholes of arbitrary shape,''
Phys. Rev. D \textbf{81}, 124036 (2010)
doi:10.1103/PhysRevD.81.124036
[arXiv:1004.1284 [hep-th]].

\bibitem{Detweiler123}
S.~Detweiler,
``On resonant oscillations of a rapidly rotating black hole,''
Proc. Roy. Soc. Lond. A \textbf{352}, 381-395 (1977)
doi:10.1098/rspa.1977.0005

\bibitem{Sun:1988tz}
Y.~Sun and R.~H.~Price,
``Excitation of Quasinormal Ringing of a Schwarzschild Black Hole,''
Phys. Rev. D \textbf{38}, 1040-1052 (1988)
doi:10.1103/PhysRevD.38.1040

\bibitem{Davidson:1981zd}
A.~Davidson and K.~C.~Wali,
``Minimal flavor unification via multigenerational Peccei-Quinn symmetry,''
Phys. Rev. Lett. \textbf{48}, 11 (1982)
doi:10.1103/PhysRevLett.48.11

\bibitem{Wilczek:1982rv}
F.~Wilczek, 
``Axions and Family Symmetry Breaking,''
Phys. Rev. Lett. \textbf{49}, 1549-1552 (1982)
doi:10.1103/PhysRevLett.49.1549

\bibitem{Gelmini:1982zz}
G.~B.~Gelmini, S.~Nussinov and T.~Yanagida,
``Does Nature Like Nambu-Goldstone Bosons?,''
Nucl. Phys. B \textbf{219}, 31-40 (1983)
doi:10.1016/0550-3213(83)90426-1

\bibitem{Banks:2010zn}
T.~Banks and N.~Seiberg,
``Symmetries and Strings in Field Theory and Gravity,''
Phys. Rev. D \textbf{83}, 084019 (2011)
doi:10.1103/PhysRevD.83.084019
[arXiv:1011.5120 [hep-th]].

\bibitem{Linde:2005ht}
A.~D.~Linde,
``Particle physics and inflationary cosmology,''
Contemp. Concepts Phys. \textbf{5}, 1-362 (1990)
[arXiv:hep-th/0503203 [hep-th]].

\bibitem{Zel:1967}
Y.~B.~Zeldovich and I.~D.~Novikov,
``The hypothesis of cores retarded during expansion and the hot cosmological model,"
Sov. Astron. Lett. \textbf{10}, 602 (1967)

\bibitem{Hawking:1971ei}
S.~Hawking,
``Gravitationally collapsed objects of very low mass,''
Mon. Not. Roy. Astron. Soc. \textbf{152}, 75 (1971)

\bibitem{Carr:1974nx}
B.~J.~Carr and S.~W.~Hawking,
``Black holes in the early Universe,''
Mon. Not. Roy. Astron. Soc. \textbf{168}, 399-415 (1974)

\bibitem{Stojkovic:2004hz}
D.~Stojkovic and K.~Freese,
``A Black hole solution to the cosmological monopole problem,''
Phys. Lett. B \textbf{606}, 251-257 (2005)
doi:10.1016/j.physletb.2004.12.019
[arXiv:hep-ph/0403248 [hep-ph]].

\bibitem{Vilenkin:2000jqa}
A.~Vilenkin and E.~P.~S.~Shellard,
``Cosmic Strings and Other Topological Defects,''
Cambridge University Press (2000)

\bibitem{Pen:1993nx}
U.~L.~Pen, D.~N.~Spergel and N.~Turok,
``Cosmic structure formation and microwave anisotropies from global field ordering,''
Phys. Rev. D \textbf{49}, 692-729 (1994)
doi:10.1103/PhysRevD.49.692

\bibitem{Yamaguchi:2001rf}
M.~Yamaguchi,
``Cosmological evolution of global monopoles,''
Phys. Rev. D \textbf{64}, 081301 (2001)
doi:10.1103/PhysRevD.64.081301
[arXiv:hep-ph/0103130 [hep-ph]].

\bibitem{Kibble:1984hp}
T.~W.~B.~Kibble,
``Evolution of a system of cosmic strings,''
Nucl. Phys. B \textbf{252}, 227 (1985)
doi:10.1016/0550-3213(85)90596-6

\bibitem{Albrecht:1984xv}
A.~Albrecht and N.~Turok,
``Evolution of Cosmic Strings,''
Phys. Rev. Lett. \textbf{54}, 1868-1871 (1985)
doi:10.1103/PhysRevLett.54.1868

\bibitem{Bennett:1985qt}
D.~P.~Bennett,
``The evolution of cosmic strings,''
Phys. Rev. D \textbf{33}, 872 (1986)
doi:10.1103/PhysRevD.33.872

\bibitem{Bennett:1986zn}
D.~P.~Bennett,
``Evolution of cosmic strings. 2.,''
Phys. Rev. D \textbf{34}, 3592 (1986)
doi:10.1103/PhysRevD.34.3592

\bibitem{Bennett:1987vf}
D.~P.~Bennett and F.~R.~Bouchet,
``Evidence for a Scaling Solution in Cosmic String Evolution,''
Phys. Rev. Lett. \textbf{60}, 257 (1988)
doi:10.1103/PhysRevLett.60.257

\bibitem{Albrecht:1989mk}
A.~Albrecht and N.~Turok,
``Evolution of Cosmic String Networks,''
Phys. Rev. D \textbf{40}, 973-1001 (1989)
doi:10.1103/PhysRevD.40.973

\bibitem{Allen:1990tv}
B.~Allen and E.~P.~S.~Shellard,
``Cosmic string evolution: a numerical simulation,''
Phys. Rev. Lett. \textbf{64}, 119-122 (1990)
doi:10.1103/PhysRevLett.64.119

\bibitem{Bennett:1989yp}
D.~P.~Bennett and F.~R.~Bouchet,
``High resolution simulations of cosmic string evolution. 1. Network evolution,''
Phys. Rev. D \textbf{41}, 2408 (1990)
doi:10.1103/PhysRevD.41.2408

\bibitem{Martins:1995tg}
C.~J.~A.~P.~Martins and E.~P.~S.~Shellard,
``String evolution with friction,''
Phys. Rev. D \textbf{53}, 575-579 (1996)
doi:10.1103/PhysRevD.53.R575
[arXiv:hep-ph/9507335 [hep-ph]].

\bibitem{Martins:1996jp}
C.~J.~A.~P.~Martins and E.~P.~S.~Shellard,
``Quantitative string evolution,''
Phys. Rev. D \textbf{54}, 2535-2556 (1996)
doi:10.1103/PhysRevD.54.2535
[arXiv:hep-ph/9602271 [hep-ph]].

\bibitem{Vincent:1996rb}
G.~R.~Vincent, M.~Hindmarsh and M.~Sakellariadou,
``Scaling and small scale structure in cosmic string networks,''
Phys. Rev. D \textbf{56}, 637-646 (1997)
doi:10.1103/PhysRevD.56.637
[arXiv:astro-ph/9612135 [astro-ph]].

\bibitem{Yamaguchi:1998gx}
M.~Yamaguchi, M.~Kawasaki and J.~Yokoyama,
``Evolution of axionic strings and spectrum of axions radiated from them,''
Phys. Rev. Lett. \textbf{82}, 4578-4581 (1999)
doi:10.1103/PhysRevLett.82.4578
[arXiv:hep-ph/9811311 [hep-ph]].

\bibitem{Yamaguchi:1999yp}
M.~Yamaguchi,
``Scaling property of the global string in the radiation dominated universe,''
Phys. Rev. D \textbf{60}, 103511 (1999)
doi:10.1103/PhysRevD.60.103511
[arXiv:hep-ph/9907506 [hep-ph]].

\bibitem{Yamaguchi:1999dy}
M.~Yamaguchi, J.~Yokoyama and M.~Kawasaki,
``Evolution of a global string network in a matter dominated universe,''
Phys. Rev. D \textbf{61}, 061301 (2000)
doi:10.1103/PhysRevD.61.061301
[arXiv:hep-ph/9910352 [hep-ph]].

\bibitem{Yamaguchi:2002zv}
M.~Yamaguchi and J.~Yokoyama,
``Lagrangian evolution of global strings,''
Phys. Rev. D \textbf{66}, 121303 (2002)
doi:10.1103/PhysRevD.66.121303
[arXiv:hep-ph/0205308 [hep-ph]].

\bibitem{Turok:1989ai}
N.~Turok,
``Global Texture as the Origin of Cosmic Structure,''
Phys. Rev. Lett. \textbf{63}, 2625 (1989)
doi:10.1103/PhysRevLett.63.2625

\bibitem{Spergel:1990ee}
D.~N.~Spergel, N.~Turok, W.~H.~Press and B.~S.~Ryden,
``Global texture as the origin of large scale structure: numerical simulations of evolution,''
Phys. Rev. D \textbf{43}, 1038-1046 (1991)
doi:10.1103/PhysRevD.43.1038

\bibitem{Turok:1990mi}
N.~Turok,
``Global texture as the origin of cosmic structure,''
Phys. Scripta T \textbf{36}, 135-141 (1991)
doi:10.1088/0031-8949/1991/T36/015

\bibitem{Bennett:1992fy}
D.~P.~Bennett and S.~H.~Rhie,
``COBE's constraints on the global monopole and texture theories of cosmic structure formation,''
Astrophys. J. Lett. \textbf{406}, L7-L10 (1993)
doi:10.1086/186773
[arXiv:hep-ph/9207244 [hep-ph]].

\bibitem{Durrer:1998rw}
R.~Durrer, M.~Kunz and A.~Melchiorri,
``Cosmic microwave background anisotropies from scaling seeds: Global defect models,''
Phys. Rev. D \textbf{59}, 123005 (1999)
doi:10.1103/PhysRevD.59.123005
[arXiv:astro-ph/9811174 [astro-ph]].

\bibitem{Bevis:2004wk}
N.~Bevis, M.~Hindmarsh and M.~Kunz,
``WMAP constraints on inflationary models with global defects,''
Phys. Rev. D \textbf{70}, 043508 (2004)
doi:10.1103/PhysRevD.70.043508
[arXiv:astro-ph/0403029 [astro-ph]].

\bibitem{Lopez-Eiguren:2017dmc}
A.~Lopez-Eiguren, J.~Lizarraga, M.~Hindmarsh and J.~Urrestilla,
``Cosmic Microwave Background constraints for global strings and global monopoles,''
JCAP \textbf{07}, 026 (2017)
doi:10.1088/1475-7516/2017/07/026
[arXiv:1705.04154 [astro-ph.CO]].

\bibitem{Hiscock:1990ev}
W.~A.~Hiscock,
``Astrophysical bounds on global monopoles,''
Phys. Rev. Lett. \textbf{64}, 344-347 (1990)
doi:10.1103/PhysRevLett.64.344

\bibitem{Herdeiro:2015waa}
C.~A.~R.~Herdeiro and E.~Radu,
``Asymptotically flat black holes with scalar hair: a review,''
Int. J. Mod. Phys. D \textbf{24}, no.09, 1542014 (2015)
doi:10.1142/S0218271815420146
[arXiv:1504.08209 [gr-qc]].

\bibitem{Ortiz:1991eu}
M.~E.~Ortiz,
``Curved space magnetic monopoles,''
Phys. Rev. D \textbf{45}, 2586-2589 (1992)
doi:10.1103/PhysRevD.45.R2586

\bibitem{Lee:1991vy}
K.~M.~Lee, V.~P.~Nair and E.~J.~Weinberg,
``Black holes in magnetic monopoles,''
Phys. Rev. D \textbf{45}, 2751-2761 (1992)
doi:10.1103/PhysRevD.45.2751
[arXiv:hep-th/9112008 [hep-th]].

\bibitem{Breitenlohner:1991aa}
P.~Breitenlohner, P.~Forgacs and D.~Maison,
``Gravitating monopole solutions,''
Nucl. Phys. B \textbf{383}, 357-376 (1992)
doi:10.1016/0550-3213(92)90682-2

\bibitem{Breitenlohner:1994di}
P.~Breitenlohner, P.~Forgacs and D.~Maison,
``Gravitating monopole solutions. 2,''
Nucl. Phys. B \textbf{442}, 126-156 (1995)
doi:10.1016/S0550-3213(95)00100-X
[arXiv:gr-qc/9412039 [gr-qc]].

\bibitem{Aichelburg:1992st}
P.~C.~Aichelburg and P.~Bizon,
``Magnetically charged black holes and their stability,''
Phys. Rev. D \textbf{48}, 607-615 (1993)
doi:10.1103/PhysRevD.48.607
[arXiv:gr-qc/9212009 [gr-qc]].

\bibitem{Lee:1994sk}
K.~M.~Lee and E.~J.~Weinberg,
``Nontopological magnetic monopoles and new magnetically charged black holes,''
Phys. Rev. Lett. \textbf{73}, 1203-1206 (1994)
doi:10.1103/PhysRevLett.73.1203
[arXiv:hep-th/9406021 [hep-th]].

\bibitem{Breitenlohner:1997hm}
P.~Breitenlohner, G.~V.~Lavrelashvili and D.~Maison,
``Mass inflation and chaotic behavior inside hairy black holes,''
Nucl. Phys. B \textbf{524}, 427-443 (1998)
doi:10.1016/S0550-3213(98)00177-1
[arXiv:gr-qc/9703047 [gr-qc]].

\bibitem{Galtsov:1997lak}
D.~V.~Gal'tsov, E.~E.~Donets and M.~Y.~Zotov,
``Singularities inside nonAbelian black holes,''
JETP Lett. \textbf{65}, 895-901 (1997)
doi:10.1134/1.567446
[arXiv:gr-qc/9706063 [gr-qc]].

\bibitem{Breitenlohner:1997ud}
P.~Breitenlohner, G.~V.~Lavrelashvili and D.~Maison,
``NonAbelian black holes: The Inside story,''
Annals Israel Phys. Soc. \textbf{13}, 172 (1997)
[arXiv:gr-qc/9708036 [gr-qc]].

\bibitem{Hartmann:2001ic}
B.~Hartmann, B.~Kleihaus and J.~Kunz,
``Axially symmetric monopoles and black holes in Einstein-Yang-Mills-Higgs theory,''
Phys. Rev. D \textbf{65}, 024027 (2002)
doi:10.1103/PhysRevD.65.024027
[arXiv:hep-th/0108129 [hep-th]].

\bibitem{Brihaye:2002km}
Y.~Brihaye and B.~Hartmann,
``SU(5) monopoles and nonAbelian black holes,''
Phys. Rev. D \textbf{67}, 044001 (2003)
doi:10.1103/PhysRevD.67.044001
[arXiv:hep-th/0211066 [hep-th]].

\bibitem{Kleihaus:2004gm}
B.~Kleihaus, J.~Kunz and F.~Navarro-Lerida,
``Rotating black holes with monopole hair,''
Phys. Lett. B \textbf{599}, 294-300 (2004)
doi:10.1016/j.physletb.2004.08.046
[arXiv:gr-qc/0406094 [gr-qc]].

\bibitem{Maldacena:2020skw}
J.~Maldacena,
``Comments on magnetic black holes,''
[arXiv:2004.06084 [hep-th]].

\bibitem{Ghosh:2020tdu}
D.~Ghosh, A.~Thalapillil and F.~Ullah,
``Astrophysical hints for magnetic black holes,''
[arXiv:2009.03363 [hep-ph]].

\bibitem{Bai:2020spd}
Y.~Bai, J.~Berger, M.~Korwar and N.~Orlofsky,
``Phenomenology of Magnetic Black Holes with Electroweak-Symmetric Coronas,''
[arXiv:2007.03703 [hep-ph]].

\bibitem{Dvali:2016mur}
G.~Dvali and A.~Gu{\ss}mann,
``Skyrmion Black Hole Hair: Conservation of Baryon Number by Black Holes and Observable Manifestations,''
Nucl. Phys. B \textbf{913}, 1001-1036 (2016)
doi:10.1016/j.nuclphysb.2016.10.017
[arXiv:1605.00543 [hep-th]].

\bibitem{Gussmann:2016mkp}
A.~Gu{\ss}mann,
``Scattering of massless scalar waves by magnetically charged black holes in Einstein–Yang–Mills–Higgs theory,''
Class. Quant. Grav. \textbf{34}, no.6, 065007 (2017)
doi:10.1088/1361-6382/aa606c
[arXiv:1608.00552 [hep-th]].

\bibitem{Andersson:19}
N.~Andersson, M.~E.~Araujo and B.~F.~Schutz,
``The phase-integral method and black hole normal modes,''
Class. Quant. Grav. \textbf{10}, 735-755 (1993)

\bibitem{Andersson:1992scr}
N.~Andersson and S.~Linnæus,
``Quasinormal modes of a Schwarzschild black hole: Improved phase-integral treatment,''
Phys. Rev. D \textbf{46}, no.10, 4179 (1992)
doi:10.1103/PhysRevD.46.4179

\bibitem{Froeman:1992gp}
N.~Froeman, P.~O.~Froeman, N.~Andersson and A.~Hoekback,
``Black hole normal modes: Phase integral treatment,''
Phys. Rev. D \textbf{45}, 2609-2616 (1992)
doi:10.1103/PhysRevD.45.2609



\end{thebibliography}
\end{document}